\documentclass[apj]{emulateapj}

\def\deg{${}^\circ$}
\def\min{${}^{\prime}$}
\def\Sec{${}^{\prime\prime}$\llap{.}}
\def\Timesec{${}^s$\llap{.}}

\shorttitle{The Tucana dSph galaxy}
\shortauthors{Monelli et al.}

\begin{document}

\title{The ACS LCID project. VI. The SFH of the Tucana dSph \\ 
and the relative ages of the isolated dSph galaxies
\altaffilmark{1}}

\author{M. Monelli\altaffilmark{2,3},
    C. Gallart\altaffilmark{2,3},
    S.L. Hidalgo\altaffilmark{2,3},
    A. Aparicio\altaffilmark{2,3},
    E.D. Skillman\altaffilmark{4},
    A.A. Cole\altaffilmark{5},
    D.R. Weisz\altaffilmark{4},
    L. Mayer\altaffilmark{6,7},
    E.J. Bernard\altaffilmark{8},
    S. Cassisi\altaffilmark{9},
    A.E. Dolphin\altaffilmark{10},
    I. Drozdovsky\altaffilmark{2,3,12},
    P.B. Stetson\altaffilmark{11}
    }
    
\altaffiltext{1}{Based on observations made with the NASA/ESA {\it Hubble Space
   Telescope}, obtained at the Space Telescope Science Institute, which is
    operated by the Association of Universities for Research in Astronomy,
    Inc., under NASA contract NAS5-26555. These observations are associated
    with program 10505.}
\altaffiltext{2}{Instituto de Astrof\'{i}sica de Canarias, La Laguna, Tenerife,
    Spain; ebernard@iac.es, monelli@iac.es, carme@iac.es, dio@iac.es,
    antapaj@iac.es, slhidalgo@iac.es.}
\altaffiltext{3}{Departamento de Astrof\'{i}sica, Universidad de La Laguna,
    Tenerife, Spain.}
\altaffiltext{4}{Department of Astronomy, University of Minnesota,
    Minneapolis, USA; skillman@astro.umn.edu.}
\altaffiltext{5}{School of Mathematics \& Physics, University of Tasmania,
    Hobart, Tasmania, Australia; andrew.cole@utas.edu.au}
\altaffiltext{6}{Institut f\"ur Theoretische Physik, University of Zurich,
    Z\"urich, Switzerland; lucio@physik.unizh.ch}
\altaffiltext{7}{Department of Physics, Institut f\"ur Astronomie,
    ETH Z\"urich, Z\"urich, Switzerland; lucio@phys.ethz.ch.}
\altaffiltext{8}{Institute for Astronomy, University of Edinburgh, Royal 
    Observatory, Blackford Hill, Edinburgh EH9 3HJ, UK; ejb@roe.ac.uk}
\altaffiltext{9}{INAF-Osservatorio Astronomico di Collurania,
    Teramo, Italy; cassisi@oa-teramo.inaf.it.}
\altaffiltext{10}{Raytheon; 1151 E. Hermans Rd., Tucson, AZ 85706, USA}
\altaffiltext{11}{Dominion Astrophysical Observatory, Herzberg Institute of
    Astrophysics, National Research Council, 5071 West Saanish Road, Victoria,
    British Columbia V9E 2E7, Canada; peter.stetson@nrc-cnrc.gc.ca.}
\altaffiltext{12}{Astronomical Institute, St. Petersburg State University,
    St. Petersburg, Russia}

\begin{abstract}
We present a detailed study of the star formation history (SFH) of the Tucana
dwarf spheroidal galaxy. High quality, deep HST/ACS  data, collected in the framework of
the LCID project, allowed us to obtain the  deepest color-magnitude diagram
to date, reaching the old main sequence turnoff ($F814 \sim 29$) with
good photometric accuracy. Our analysis, based on
three different SFH codes, shows that Tucana is an old and metal-poor
stellar system, which experienced a strong initial burst of star formation
at a very early epoch ($\simeq$ 13~Gyr ago) which lasted a maximum of 1~Gyr
(sigma value). We are not able to unambiguously answer the question of
whether most star formation in Tucana occurred before or after the end of
the reionization era, and we analyze alternative scenarios that may explain
the transformation of Tucana from a gas-rich galaxy into a dSph.
Current measurements of its radial velocity do not preclude that Tucana may
have crossed the inner regions of the Local Group once, and so 
gas stripping by ram pressure and tides due to a close interaction cannot be ruled
out. A single pericenter passage would generate insufficient tidal heating
to turn an originally disky dwarf into a true dSph; however, this possibility
would be consistent with the observed residual rotation in Tucana.
On the other hand, the high star formation rate measured at early
times may have injected enough energy into the interstellar medium 
to blow out a significant fraction of the initial gas content. Gas that is
heated but not blown out would also be more easily stripped via ram
pressure. We compare the SFH inferred for Tucana with that of Cetus, the other
isolated LG dSph galaxy in the LCID sample. We show that the formation time
of the bulk of star formation in Cetus is clearly delayed with respect to
that of Tucana. This reinforces the conclusion of Monelli et al.\ (2010)
that Cetus formed the vast majority of its stars after the end of the
reionization era implying, therefore, that small dwarf galaxies are
not necessarily strongly affected by reionization, in agreement with many
state-of-the-art cosmological models. 
\end{abstract}


\section{Introduction}\label{sec:intro}

Nearby galaxies in the Local Group (LG) present a variety of different 
properties in terms of stellar populations, gas and metal content, 
morphological type, and mass. Understanding their differences and similarities and
how these evolved with time can bring important insights into our knowledge
of the formation and evolution of the LG, and of galaxies in general. The 
recovery of their full star formation history (SFH) plays a key role, being among
the most powerful techniques to dig into the mechanisms that drove the
evolution of stellar systems. Precise estimates of the epochs of star
formation and their duration give direct information to compare the
observed properties with predictions. This is particularly important
for investigating the first stages of the life of galaxies, because the
properties of the stellar populations during the first few~Gyr can be used
to trace the impact of a variety of different physical mechanisms. For
example, a delay in the onset of the star formation, or a truncation of the
SFH at a very old epoch can possibly be used to infer the role of 
cosmic reionization \citep[e.g.,][]{ikeuchi86, rees86, efstathiou92, babul92,
chiba94, quinn96, thoul96, kepner97, barkana99, bullock00, tassis03, ricotti05, 
gnedin06, okamoto09, sawala10, mayer10}. On the other hand, an episodic 
rather than continuous SFH might trace interactions of dwarf galaxies 
in their orbit around bigger systems \citep{mayer01}, or a possible cycle 
related to changes in the interstellar medium linked to supernova explosions
\citep{carraro01, stinson07, valcke08}. What is crucial in these kinds of
analysis is the capability to precisely and accurately recover how these
systems formed stars as a function of time, and to characterize their
chemical evolution.

Within this general context, we selected a sample of isolated dwarf
galaxy members of the LG, with the aim of deriving quantitative SFHs over
their entire lifetime. The main objective of the LCID project ({\itshape
Local Cosmology from Isolated Dwarfs}\footnotemark[14]) is to study the
details of the evolution of nearby isolated galaxies, compare their
properties and help constrain cosmological and galaxy formation and
evolution models.
\footnotetext[14]{http://www.iac.es/project/LCID}
A general description of the project, and an overview of the results,
are presented in Gallart et al. (in prep). This paper is focused
on the study of one the most isolated dSph galaxies in the LG, Tucana,
with particular emphasis on the comparison its properties with those of
the Cetus dSph \citep{monellicetus}.

Understanding the currently isolated nature of Tucana and Cetus is challenging 
in the context of galaxy-formation theories. Models have been proposed
suggesting that dSph galaxies originate from small, initially gas-rich 
galaxies that lost their gas through tidal and ram-pressure interactions 
with their large host galaxies \citep[see ][for a review]{mayer10}.
Before the LCID project, however, the existing observations
of Cetus and Tucana did not allow us to draw definite conclusions on the
exact nature of these galaxies, particularly regarding the question of how
extended in time their SFHs were. In \citet{monellicetus} and in this
paper, we show that Cetus and Tucana are similar to the oldest Milky Way
dSph satellites such as Draco, Ursa Minor, Sculptor and Sextans, and therefore
they do not follow the morphology-density relation observed among
Milky Way dSph satellites (e.g. \citealt{vandenbergh99}). Their mere 
existence, therefore, imposes new constraints on dSph galaxy-formation models.

The first mention of the Tucana dSph appears in the Southern Galaxy Catalogue 
published by \citet{corwin85}, but it was later "rediscovered" by 
\citet{lavery90}. The first CMD was presented 
in \citet{lavery92}, based on Anglo-Australian 3.9m $V$, $I$ images. Tucana 
appeared as an elongated spheroid (e=0.5), and was classified as a dE5.
Although their photometry was quite shallow, \citet{lavery92} could give the 
first estimates of the distance ($(m-M)$ $<$ 24.75), metal content ([Fe/H] = $-$1.9)
and luminosity (M$_V$ = $-$9.5). Tucana was definitively recognized as a 
member of the LG, with the peculiarity of being the first isolated 
dSph, not linked with either the Milky Way or M31. \citet{lavery92} also 
inferred that Tucana is a predominantly old system, based on the lack of 
young bright blue objects.

All these preliminary findings were later substantially confirmed by 
deeper data. \citet{castellani96} presented $V$, $I$ photometry reaching 
$V \sim$ 26 mag, which allowed them to estimate the distance ($(m-M)_I$
= 24.72 $\pm$ 0.20) based on the tip of the red giant branch (RGB), and a
mean metallicity of [Fe/H] = $-$1.56, obtained comparing the Tucana RGB
with the ridge line of Galactic globular clusters (GCs) from \citet{dacosta90}.
Interestingly enough, they found that the color width of the RGB was 
larger than expected on the basis of the photometric errors. They 
interpreted this as a signature of a ''consistent spread'' in the 
metallicity of the Tucana stars.

A different conclusion was reached by \citet{saviane96}, who presented 
$V$, $I$ data collected with the EFOSC camera mounted on the 2.2m ESO 
telescope. A comparison with the same database from \citet{dacosta90} 
yielded a mean metallicity [Fe/H]$\sim -$1.80, and ``no convincing 
indication of an abundance spread.'' An interesting point raised by 
\citet{saviane96} concerns the absolute dimension of Tucana. Using an 
exponential profile, they estimated a core radius of 166 pc, similar
to that of the small Milky Way and M31 satellites (Mateo 1998).

Another very interesting characteristic of Tucana is the strong horizontal 
branch morphology gradient first detected by \citet{harbeck01}, together
with a double peak in the color distribution of the RGB stars. Based on 
these two hints, and on the likely absence of intermediate age stars 
(supported by the non-detection of C stars, \citealt{battinelli00}), 
\citet{harbeck01} suggested that Tucana experienced a very fast 
self-enrichment. This possibility was strongly supported by
\citet{bernard08} on the basis of the properties of a sample of $\sim
$400 RR Lyrae stars. They detected two different sub-populations of RR
Lyraes having distinct spatial distributions and different mean luminosities, and
following two period-amplitude relations with different slopes. This
supports a scenario with two different sub-populations older than 10~Gyr,
the more centrally concentrated one being slightly younger and
slightly more metal rich than the other.

The paper is organized as follows. \S \ref{sec:obse} summarizes the 
observations and data reduction strategy. In sect. \S \ref{sec:cmd} we present 
the CMD, while the derivation of the SFH is extensively discussed in \S 
\ref{sec:sfh}. In \S \ref{sec:cetucana} we compare the properties 
of the two isolated dSphs Cetus and Tucana. Discussion and conclusions
are presented in \S \ref{sec:discussion} and \S \ref{sec:summa}.

\section{Observations and data reduction}\label{sec:obse}

The data presented in this paper were collected with the ACS 
camera aboard the HST \citep{ford98}, as part of the project 
{\itshape The onset of star formation in the universe: constraints 
from nearby isolated dwarf galaxies} (PID 10505, PI C. Gallart).


\begin{figure*}
\epsscale{.80}
\plotone{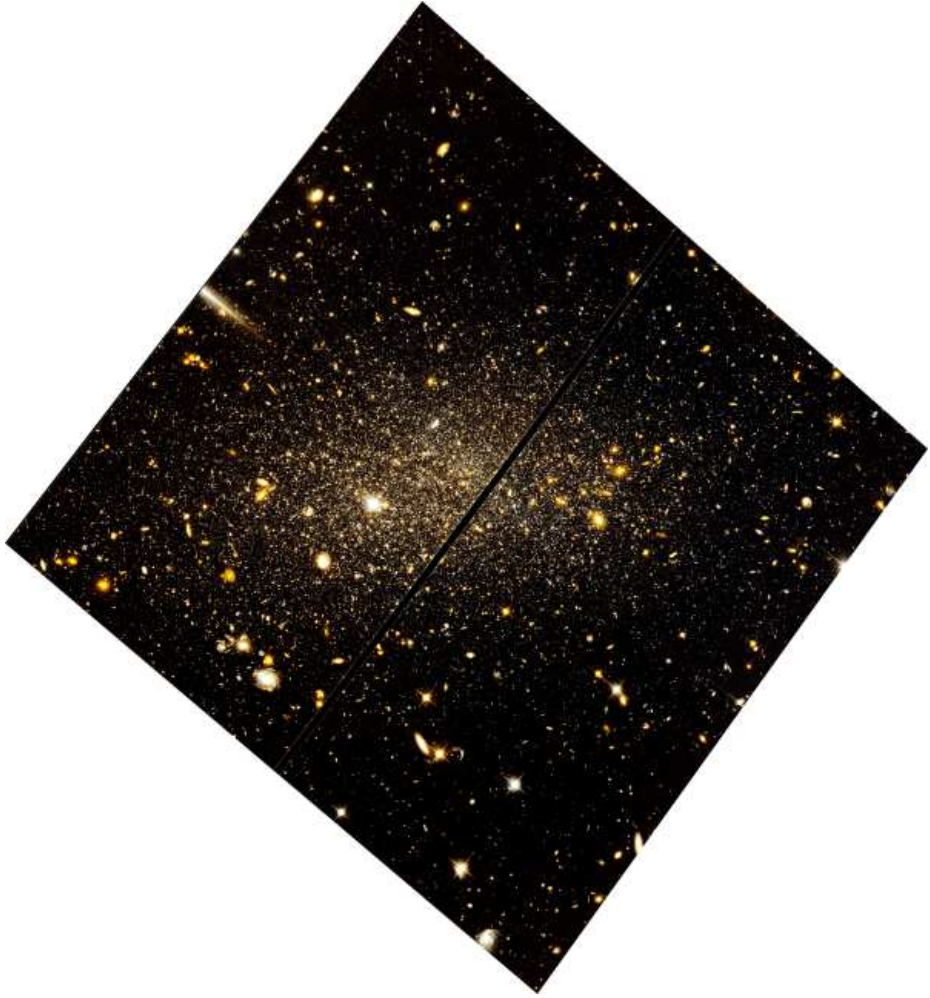}
\caption{Stacked, drizzled  color image of the Tucana field. North is up 
and East is left.  The field of view is $\sim$3\min4$\times$3\min4. 
The image shows a clear gradient in the number of stars when moving from
the center to the outskirts. A sizable number of background galaxies is visible
as well.
\label{fig:map}}
\end{figure*}
%

The 32 orbits allocated to Tucana were executed between April 25
and 30, 2006. The observations were split into visits, each of them
including  two orbits. One $F475W$ and one $F814W$ image were collected
during each orbit,  with exposure times slightly different among the two
orbits of the same visit: 1,070$\,$s and 957$\,$s in the first orbit, 1,090$\,$s and
979$\,$s in the second one.  Summarizing, the total observing time on Tucana
was 34,560$\,$s in $F475W$ and 30,976$\,$s in $F814W$. Note that, despite
the fact that Tucana is the most distant galaxy in the LCID sample, the 
exposure times adopted are the shortest. This choice was forced due to 
the constraints imposed by the South Atlantic Anomaly.


\begin{figure*}
\epsscale{1.}
\plotone{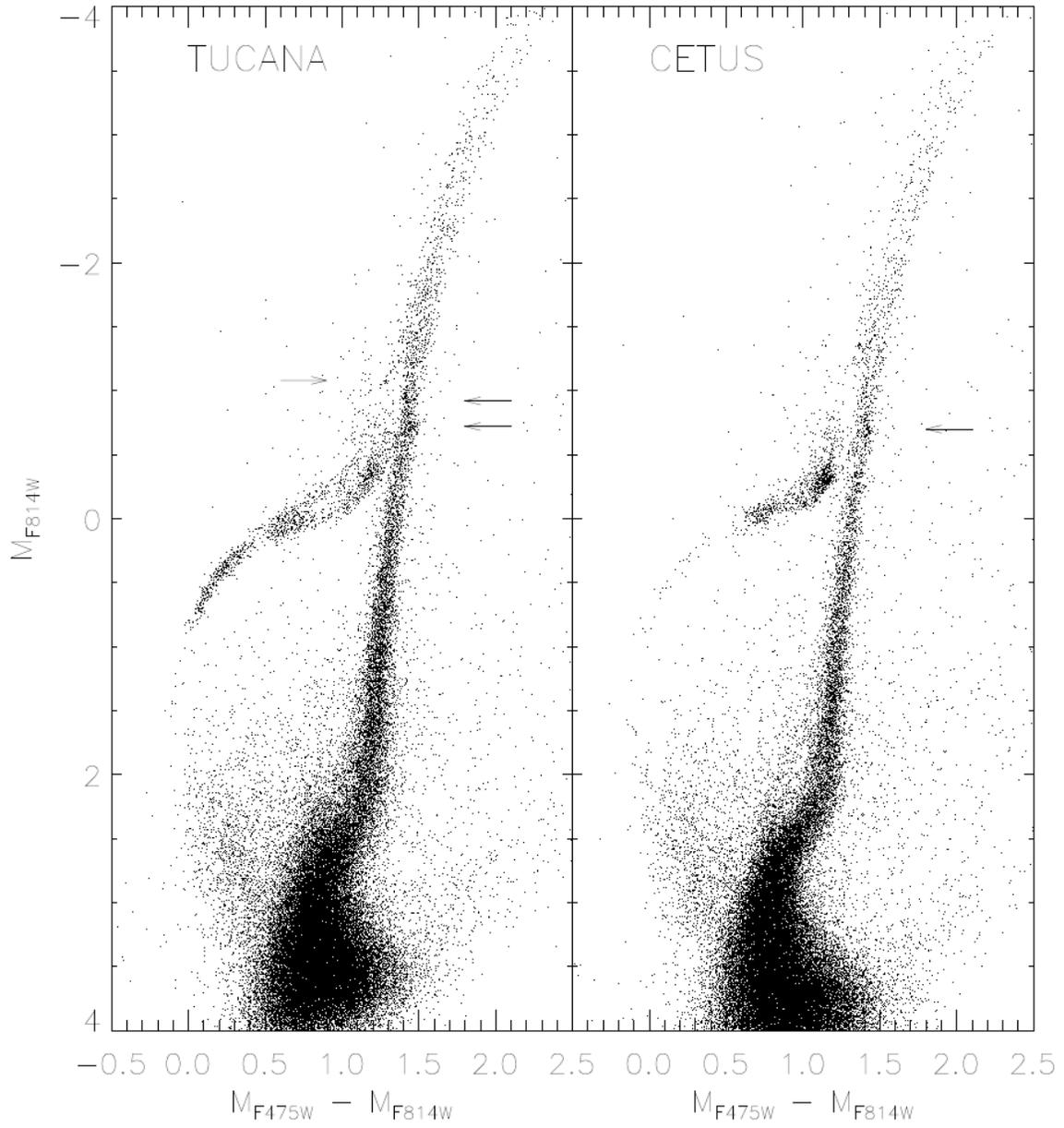}
\vspace{2.5truecm}
\caption{CMD of Tucana and Cetus obtained with DAOPHOT.  Note that the color of the 
bluest HB stars is similar in both galaxies, but the number of blue HB stars is 
significantly higher in Tucana than in Cetus. The RGB bumps are also obvious features 
in both galaxies as indicated by the arrows on the right side of the RGBs.
The arrow on the left side of the Tucana RGB mark the AGB clump. \citep{monelli10b}
\label{fig:cetucana_cmd}}
\end{figure*}


The pointing ($\alpha$ = 22${}^h\,$41${}^m\,$48\Timesec43, $\delta$ = --64\deg$\,$25\min$\,$15\Sec7)
was centered on the galaxy. Figure \ref{fig:map} shows a drizzled, stacked image.
The parallel WFPC2 field already presented in \citet{bernard09} was 
intended to sample the halo of Tucana along its major axis, at the position
(22${}^h\,$40${}^m\,$52\Timesec92;  --64\deg$\,$24\min$\,$28\Sec9). The CMD of this 
field, located 6\min from the Tucana center, does not show any convincing 
evidence that stars of Tucana are still present at this distance, since no 
RR Lyrae stars were detected.

	\subsection{Data reduction}

The details of the data reduction strategy, common to all the LCID galaxies,
have been presented in \citet{monellicetus}, 
so here we just recall the main points. We performed two parallel and independent 
photometric reductions, using the DAOPHOT/ALLFRAME \citep{alf} and DOLPHOT
packages \citep{dolphot}. We applied both codes to the original \_FLT images, working
on the two chips separately. In the case of DAOPHOT, we modeled 
individual PSFs for each frame, using bright and isolated stars in the field.
Typically, 400 stars and a Moffat function with index $\beta=1.5$ were used. 

The DOLPHOT code was applied using the ACS module and following the 
recommended photometry recipe provided in the manual for version 1.0.3.  

Individual photometry catalogs for each frame were calibrated.
We applied the calibration provided by \citet{sirianni05} identically to both sets 
of photometry. The final CMD is presented in Fig. \ref{fig:cetucana_cmd}  and 
discussed in section \ref{sec:cmd}.

	\subsection{Completeness tests}

The completeness tests were performed by adding synthetic stars to the individual
images, and repeating the photometric analysis identically as for the real data.

In the case of the DAOPHOT photometry the list of injected stars was
created  following the prescriptions introduced by \citet{gallart96}, using
a synthetic CMD created  with IAC-star \citep{iacstar}. This was built
adopting a constant star formation rate  between 0 and 15~Gyr, and a 
metallicity distribution following a flat distribution in the range 0.0001 $ <
Z < $ 0.005. Approximately 1,500,000 stars were simulated  in each of
the two chips, in iterations of $\sim$ 50,000 stars.

In the case of DOLPHOT, we followed the method described in
\citet{holtzman06}, simulating one star at a time and covering the range 
$-1 < M_{F475W} - M_{814W} < 5$ mag, $-7 < M_{F475W} < 8$ mag. The density
of injected stars is higher in the color-magnitude range where most of
the observed stars are found. A total of 140,000 stars were simulated.
We found that, at the level of the TO, the completeness is of the order 
of $80\%$.

	\subsection{Comparison of the two photometry setss}

We detect a small zero point offset for the brightest stars in the $m_{F475W}$ 
($m_{F475W,DAO}-m_{F475W,DOL}= +0.019$ mag, for $m_{F475W} <
25$) and in the $m_{F814W}$ 
($m_{F814W,DAO}-m_{F814W,DOL}= - 0.008$ mag,  for $m_{F814W} <
24$) bands, and a trend as a function of magnitude. As already noticed in
the case of  Cetus \citep{monellicetus}, moving toward fainter magnitudes,
DAOPHOT  tends to measure stars slightly fainter. However, since the trend
with magnitude is similar in the two bands, the color difference shows a
small residual zero point ($\sim0.032$ mag), but no dependence on the
magnitude. As already noticed in the case of Cetus \citep{monellicetus}
and LGS 3 \citep{hidalgolgs3},
and discussed  in \cite{holtzman06} in the case of the WFPC2, small
differences in the photometry are expected when using different packages.
Most importantly, we will show in section \ref{sec:sfhresults} that the
impact of the differences in the photometry on the derived SFH is
negligible.

\section{The CMD}\label{sec:cmd}

Fig. \ref{fig:cetucana_cmd}, left panel, presents the final CMD for Tucana from the DAOPHOT
photometry, calibrated to the VEGAMAG system and shifted to the absolute plane.
The right panel shows the CMD of Cetus \citep{monellicetus}. The comparison 
of the two CMDs shows striking similarities in the different evolutionary 
features. An in-depth comparison of the CMDs and the properties of these two 
galaxies is presented in \S \ref{sec:cetucana}.

The Tucana CMD presented here is the deepest obtained for this galaxy 
to date. It spans more than $\sim$8 mag,
from the  tip of the RGB down to $\approx$1.5 mag below the oldest main sequence
(MS) turn-off (TO), or $M_{F814W} \sim $4. The absence of a blue young MS
and the morphology of the old MSTO, at  $M_{F814W} \sim$ 2.5 mag, are typical
features of an old stellar system, with no strong recent or intermediate-age
star formation. In agreement with what is commonly found in stellar systems,
the plume of relatively faint objects appearing at 1.5 $< M_{F814W} <$ 3, 
0 $< (M_{F475W} - M_{F814W}) <$ 0.5 is most likely a population of blue stragglers.
The possibility of some residual star formation occurring in the last 3-4~Gyr 
is discussed in \S \ref{sec:sfhresults}. The RGB appears as a dominant feature, 
between -4 $< M_{F814W} <$ 2 mag. The width of the RGB suggests a limited spread 
in metallicity. The two over-densities located along the RGB at $M_{F814W} \sim 
-0.9, -0.7$ and $M_{F475W} - M_{F814W} \sim +1.4$ and highlighted by two arrows 
are the RGB bumps (see \citealt{monelli10b} for a detailed study of the RGB 
bump feature in several LCID galaxies). The other clear over-density at $M_{F814W} \sim 
-0.8$ and  $M_{F475W} - M_{F814W} \sim 1.2$, and joining the RGB from the 
blue side, can be identified with the AGB bump and the subsequent AGB.

\begin{figure}
\epsscale{1.2h}
\resizebox{8truecm}{7truecm}{\includegraphics[clip=true]{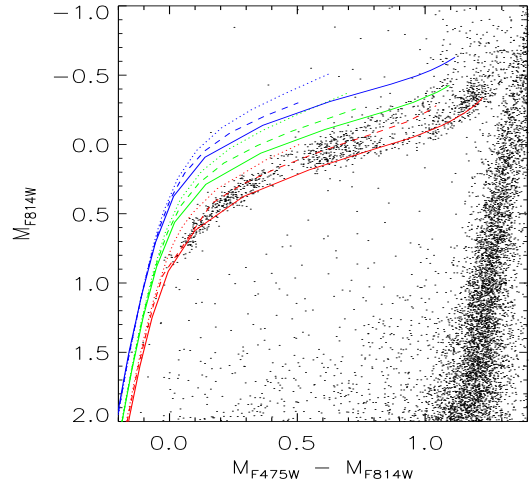}}
\caption{Tucana CMD with superimposed ZAHBs of different metallicities Z=0.0001
(dotted), 0.0003 (dashed), 0.0006 (solid) and Y= 0.245 (red), 0.30 (green), 
0.35 (blue) are shown. Only the theoretical predictions with normal helium 
give a good representation of the HB observed in Tucana.
\label{fig:helium}}
\end{figure}

The Tucana HB, as previously noted by \citet{harbeck01}, shows a complex
morphology, which can be appreciated in detail in the present, much deeper
data. It is well populated from the red to the blue side, to
$M_{F475W} - M_{F814W} \sim$ 0. A few bluer and fainter objects seem to 
continue the HB sequence, but they merge with the BSs candidates and it 
is not possible to draw firm conclusions on their nature on the basis 
of the present data.
A possible scenario to explain the strong blue component of the Tucana 
HB is that this feature is populated by a helium-enriched population.
Helium-rich stars have been invoked to explain the properties
of the most massive Galactic globular clusters which have been proved to host
more than one population \citep{piotto05, caloi05, milone08, milone10}. The 
effect of the helium enrichment is to produce, at fixed age and for $M_{F475W}
 - M_{F814W} > -0.2$, a bluer and brighter HB. Fig. \ref{fig:helium} shows the 
 Tucana CMD with selected 
ZAHBs from the BaSTI database superimposed, for metallicity Z=0.0001, 0.0003, 0.0006 and 
Y=0.245, 0.30, 0.35. The figure shows that the HB is nicely bracketed by
the canonical helium ZAHBs, and a metallicity in the range 0.0003-0.001.
Note that this is in excellent agreement with the results of the SFH, suggesting
a mean metallicity of Z = 0.0005 (see \S \ref{sec:sfhresults}). The helium-rich 
ZAHBs appear too bright to explain any significant number of Tucana HB stars.


\begin{figure*}
\epsscale{1.}
\resizebox{15truecm}{15truecm}{\includegraphics[clip=true]{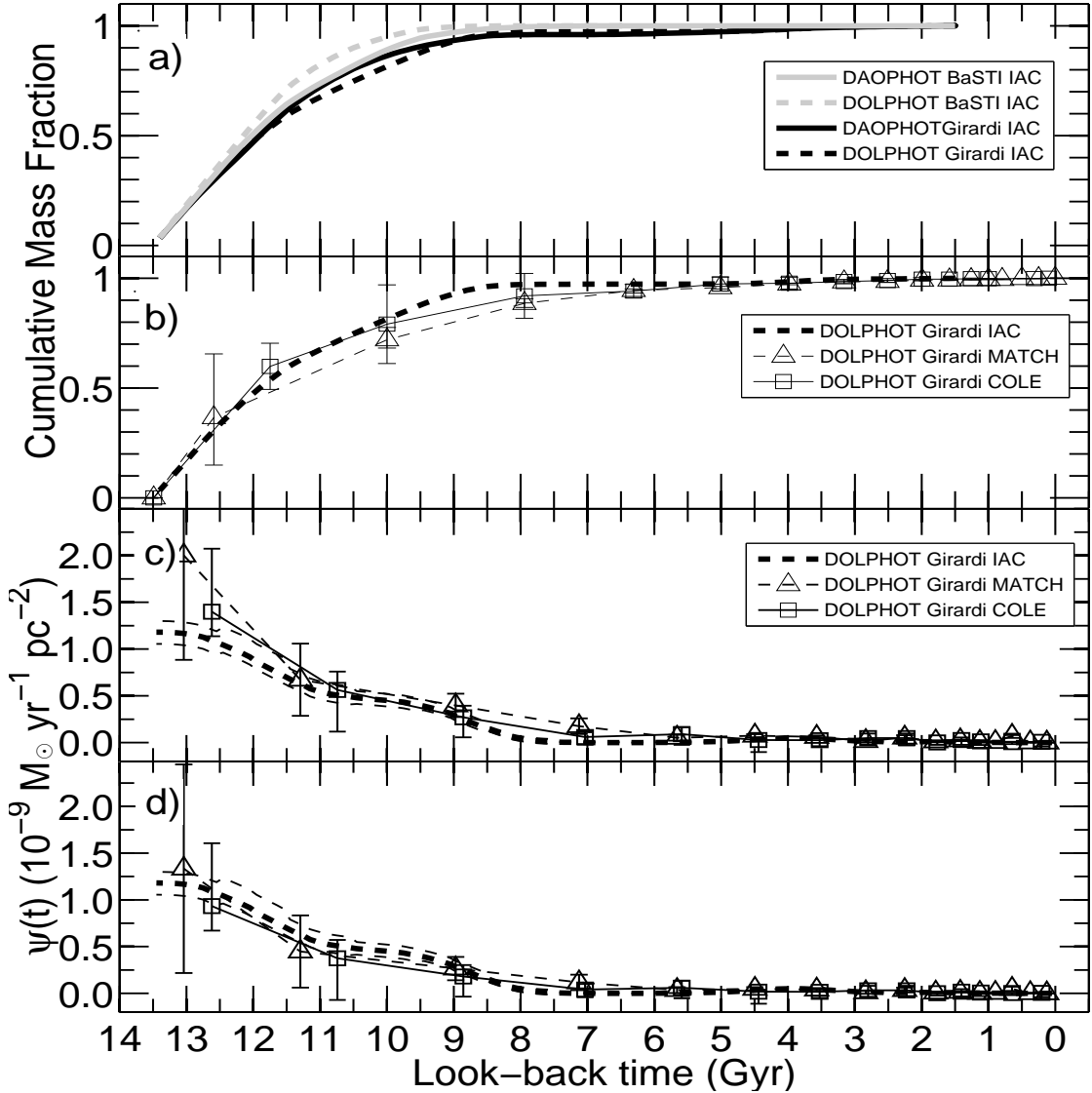}}
\caption{Summary of the results obtained with different techniques to 
derive the SFH. {\itshape Top:} Comparison of the cumulative mass fraction
derived from SFH obtained with the MinnIAC/IAC-pop method, using the BaSTI 
and Padova library and applied to the DAOPHOT and DOLPHOT photometry. 
{\itshape Middle:} Same as in the upper panel, but the comparison shows 
the results the from three different SFH codes, applied to the DOLPHOT 
photometry in combination with the Padova/Girardi {\itshape Bottom:} 
$\psi$(t) derived for the three solutions presented in the central panel.
\label{fig:external}}
\end{figure*}


\section{The star formation history of Tucana}\label{sec:sfh}

The SFH of Tucana was derived, similarly to the other galaxies of  the LCID
sample, using the two photometry sets, two stellar evolution libraries 
(BaSTI\footnotemark[15], \citealt{pietrinferni04} and
Padova/Girardi\footnotemark[16], \citealt{girardi00}) and three
different SFH codes: IAC-pop  \citep{iacpop}, MATCH \citep{dolphin02}, and
COLE \citep{skillman03}.  We performed the same kind of analysis already
presented for Cetus  \citep{monellicetus} and LGS3 \citep{hidalgolgs3}. In
particular, we used the three SFH codes together with the DOLPHOT photometry
in combination with the Padova stellar evolution library.
\footnotetext[15]{\itshape http://www.oa-teramo.inaf.it/BASTI}
\footnotetext[16]{\itshape http://pleiadi.oapd.inaf.it/} The IAC-pop
solutions were derived using both photometry sets and both libraries. 

In the following sections we summarize the fundamentals of the three SFH methods 
applied. A comparison of the results is given in \S \ref{sec:sfh_comparison},
while the Tucana SFH will be discussed in \S \ref{sec:sfhresults}.

	\subsection{The IAC method}\label{sec:method_iacpop}

The IAC method to derive the SFH of resolved stellar systems, based on the 
IAC-star/IAC-pop/MinnIAC algorithms, has been extensively presented in 
\citet{iacpop} and \citet{hidalgolgs3}. Due to the
similarities between the Tucana and Cetus CMD, we applied the same criteria
explained for Cetus in \citet{monellicetus}. The reader
is referred to that paper for more details and for the discussion of the
different choices. In summary, we have adopted the following parameters and
functions:

{\em i) Model functions- } The same synthetic CMDs containing $8 \times 10^6$ stars 
have been used as for Cetus. These were generated using the IAC-star 
simulator\footnotemark[17], using: \\
$\bullet$ the BaSTI and Padova stellar evolution libraries, respectively; \\
$\bullet$ a constant $\psi$(t) between 0 and 15~Gyr, and metallicities
uniformly distributed in the range 0.0001 $<$ Z $<$ 0.005; \\
$\bullet$ a binary fraction\footnotemark[18] $\beta$=0.4 with mass ratios 
$q \ge 0.5$; \\
$\bullet$ the IMF from \citet{kroupa01} ($N(m)~dm = m^{-\alpha}~dm$, where
$\alpha =  1.3, 2.3 $for stars with mass smaller and larger than 0.5 
M$_\odot$, respectively); \\
$\bullet$ the set of bolometric corrections from \citet{bedin05}. \\
The observational errors were simulated in the synthetic CMD using the {\itshape obsersin}
code \citep{hidalgolgs3}.

\footnotetext[17]{\itshape http://iac-star.iac.es}
\footnotetext[18]{We tested six different values, from $\beta$=0 to 1. A 
more detailed discussion can be found in \citet{monellicetus}.}

{\em ii) Simple populations and CMD sampling:} 

The initial age and metallicity bins defining the simple populations are: \\
{\itshape age:} [1.5 2 3 ...11.5 13.5]$\times 10^9$ years\\
{\itshape metallicity:} [0.1 0.3 0.5 0.7 1 1.5 2]$\times 10^{-3}$ \\

The parametrization of the CMD was performed using the same choice of 
{\itshape bundles} and box size presented in \citet{monellicetus} (see Figure 
\ref{fig:cmdbest}). Moreover, the same multiple sampling of both the observed 
and the model CMDs (``{\it dithering} approach'') has been applied, using the
{\itshape MinnIAC} code \citep{iacpop}. Twenty-four solutions  with twelve 
different samplings of the simple populations and two samplings of the positions 
of the boxes grid each, were calculated.

{\em iii) External parameters: } 

The true distance modulus, $(m-M)_0=24.77$\footnotemark[19], was estimated using the
mean magnitude of the RR Lyrae stars \citep{bernard09}. We assumed
the extinction from \citet{schlegel98}, $E(B-V) = 0.03$, corresponding to
$A_{F475W} = 0.122$ mag and  $A_{F814W} = 0.061$ mag.
\footnotetext[19]{The value adopted in the SFH derivation, 24.77, is
slightly  different from the final value of 24.74 $\pm$ 0.12  given in
\citep{bernard09}. However, such a small difference has negligible
impact on the derived SFH.}
To estimate the effect of the uncertainties
affecting the photometry and the transformation to the absolute magnitude
plane (calibration, distance, extinction) and possible model zero points,
MinnIAC introduces a grid of magnitude and color shifts to the observed
CMD. The values of the mean $\chi^2$ in each node of the grid are
analyzed to determine the best solution and the associated shift from the
adopted transformation to the absolute plane. In the process, the sets of
24 solutions were calculated 39 times, with different shifts, for a
total of 936 solutions. This computationally expensive task was performed
using the Condor workload management system\footnotemark[20] \citep{condor} 
available at the Instituto de Astrof\'isica de Canarias.
\footnotetext[20]{http://www.cs.wisc.edu/condor/}
An initial grid with 25 positions was common to the two photometry sets.
A second grid with 14 positions is used to refine the search for the best
$\chi^2$. In the case of the DAOPHOT photometry, the minimum was confirmed 
at the position ($\delta_{col}$ ; $\delta_{mag}$) = (0.0 ; 0.0),
$\chi^2_{\nu,min}$ = 2.63, while in the case of the DOLPHOT photometry 
the minimum is located at ($\delta_{col}$ ; $\delta_{mag}$) = (+0.06, 0.15), 
$\chi^2_{\nu,min}$ = 2.57. In the \S \ref{sec:sfhresults} we describe the 
details of the two solutions calculated at the $\chi^2_{\nu,min}$ position.

	\subsection{The Match method}\label{sec:method_match}

The principles of the MATCH method of measuring SFHs are presented in \citet{dolphin00} 
and \citet{dolphin02}. In this work,
to appropriately compare the synthetic and observed CMDs, we converted them 
into Hess diagrams binning by 0.10 mags in M$_{F475W}$ and 0.05 mags in M$_{F475W}$
$-$ M$_{F814W}$. For consistency, we chose parameters to create synthetic 
CMDs that closely matched those used in the IAC-pop code presented 
in this paper: \\
$\bullet$ a power law IMF with $\alpha$ $=$ 2.3 from 0.1 to 100 M$_\odot$; \\
$\bullet$ a binary fraction $\beta$ $=$ 0.40; \\
$\bullet$ a distance modulus of 24.77, A$_{F475W}$ $=$ 0.122; \\
$\bullet$ the stellar evolution libraries of \citet{marigo08},
which are the basic Padova/Girardi models with updated AGB evolutionary 
sequences.

The synthetic CMDs were populated with stars over a time range of
$\log{t}$ $=$ 6.6 to $\log{t}$ $=$ 10.20, with a uniform bin size of 0.1.
Furthermore, the program was allowed to solve for the best fit metallicity per 
time bin, drawing from a range of [M/H] = $-$2.3 to 0.1, where M canonically
represents all metals. The depth of the photometry used for the SFH 
was equal to 50\% completeness in both M$_{F475W}$ and M$_{F814W}$.

To quantify the accuracy of the resultant SFH, we tested for both 
statistical errors and modeling uncertainties. To simulate possible 
zero point discrepancies between isochrones and data, we constructed 
SFHs adopting small offsets in distance and extinction from the best
fit values. The rms scatter between those solutions is a reasonable proxy
for uncertainty in the stellar evolution models and/or photometric 
zero-points. Statistical errors were accounted for by solving 
fifty random realizations of the best fit SFH. Final error bars are calculated
summing in quadrature the statistical and systematic errors.


\begin{figure}
\epsscale{0.8}
\resizebox{8truecm}{7truecm}{\includegraphics[clip=true]{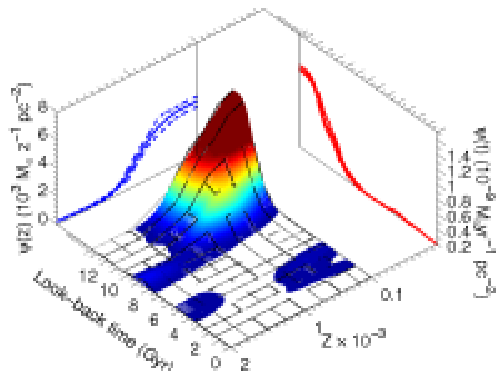}}
\resizebox{8truecm}{7truecm}{\includegraphics[clip=true]{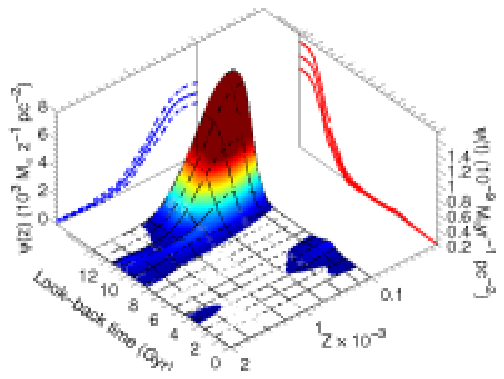}}
\caption{3-D representation showing the SFH of Tucana, derived with the 
DAOPHOT {\itshape (top)} and the DOLPHOT {\itshape (bottom)} photometry. 
Both show the average of 24 solutions calculated at the minimum $\chi^2$
in the $\delta mag$-$\delta col$ grid.
\label{fig:tucana3d}}
\end{figure}


	\subsection{The COLE method}\label{sec:method_cole}

The third method we applied to measure the SFH of Tucana is a simulated
annealing algorithm applied to our DOLPHOT photometry and the Padova/Girardi
library. The code is the same as that applied to the galaxies Leo~A 
\citep{cole07} and IC~1613 \citep{skillman03}, and details of the 
method can be found in those references.


\begin{figure*} 
\epsscale{1.}
\plotone{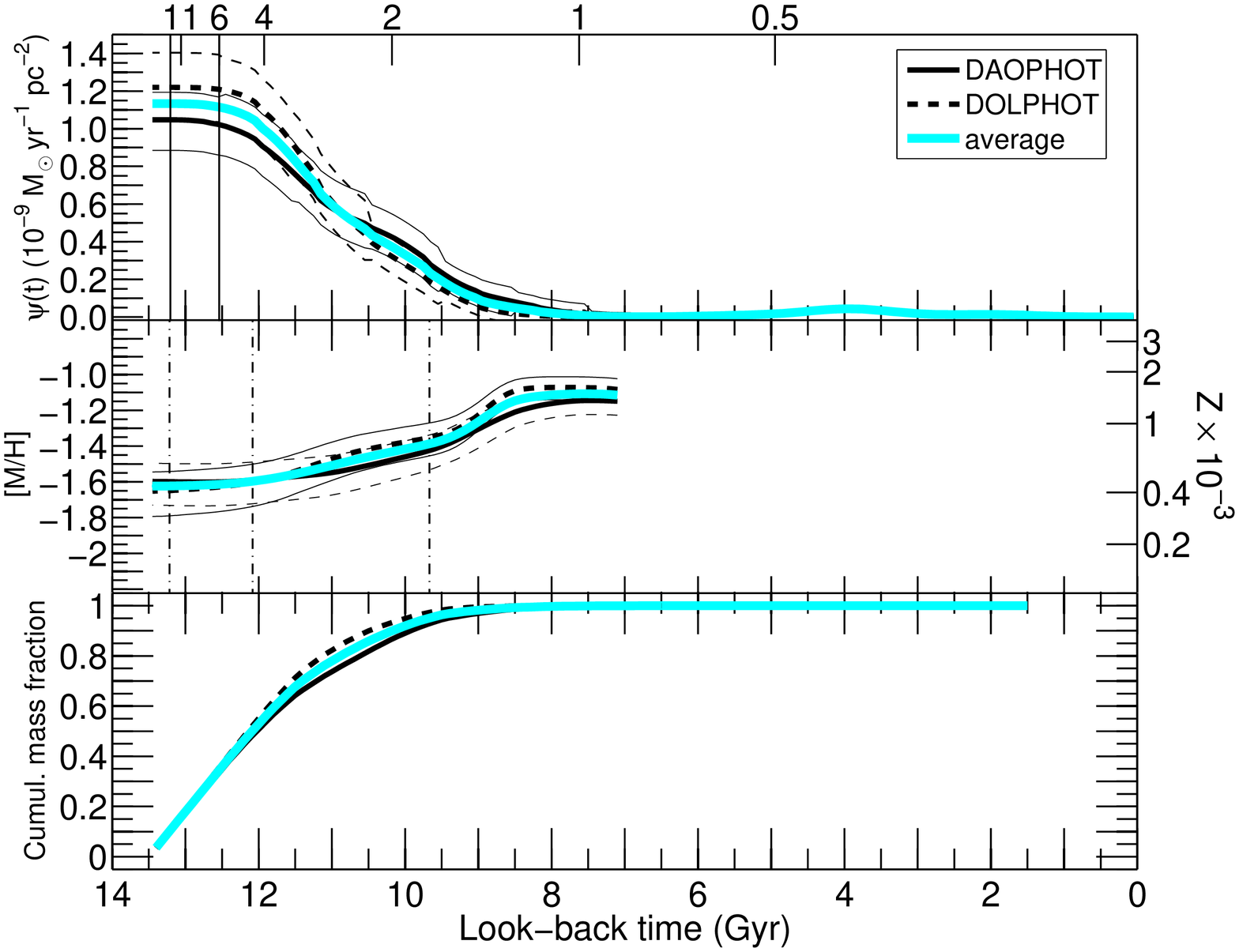}
\caption{Summary of the Tucana SFH results. The three panels show, from
top to bottom, $\psi$(t), the age-metallicity relation  and the
cumulative mass fraction. We show the results with the BaSTI  library for
both the DAOPHOT (dashed line) and the DOLPHOT (continuous) photometry, and
the average of the two (thick cyan line). The dashed  part of the cyan line
representing the average of the two sets of  photometry in the middle panel
indicates what we believe to be the presence of BSs rather than a low level of intermediate-age
and young SF. The thin lines  represent the error bars, while the vertical
lines in the upper panel  mark redshifts {\itshape z} $=$ 15 and 6---the epochs
corresponding  to the reionization. \citep{bouwens07}. The dotted-dashed lines
in the central panel mark the epochs when 10\%, 50\% and 90\% of the stellar 
mass was formed.
\label{fig:sfh_daodol}} 
\end{figure*}


In particular, the isochrone tables \citep{marigo08} for each (age, 
metallicity) pair are shifted by the appropriate distance and reddening 
values, and transformed into a discrete color-magnitude-density distribution 
$\varrho$ by convolution with: \\
$\bullet$ the initial mass function from \citet{chabrier03}; \\
$\bullet$ the results of the artificial star tests simulating observational 
uncertainties and incompleteness; \\
$\bullet$ the binary star frequency and mass ratio distribution from
the solar neighborhood, namely 35\% of stars are single, 46\% are 
parametrized as ``wide" binaries, i.e., the 
secondary mass is uncorrelated with the primary mass, and is drawn from
the same IMF as the primary, and finally 19\% of stars are parameterized
as ``close" binaries, i.e., the probability distribution function of 
the mass ratio is flat. 

The history of the 
galaxy is then divided into discrete age bins with approximately 
constant logarithmic spacing. We used 10 time bins ranging from 8.00 $\leq$ 
log(age/Gyr) $\leq$ 10.13. Isochrones evenly spaced by 0.2 dex
from $-2.3 \leq [M/H] \leq -0.9$ were used, considering only scaled-solar
abundances but with no further constraints on the age-metallicity relation.
The SFH is then modeled as the linear combination of $\psi$(t,Z) that best 
matches the data by summing the values of $\varrho$ at each point in the 
Hess diagram. The CMD was discretized into bins of dimension
0.10$\times$0.20 mag, and it is the density distribution in 
this Hess diagram that was fit to the models.  The error bars are determined
by perturbing each component of the best-fit SFH in turn and re-solving for
a new best fit with the perturbed component held fixed.  


\begin{figure*}
\epsscale{1.}
\plotone{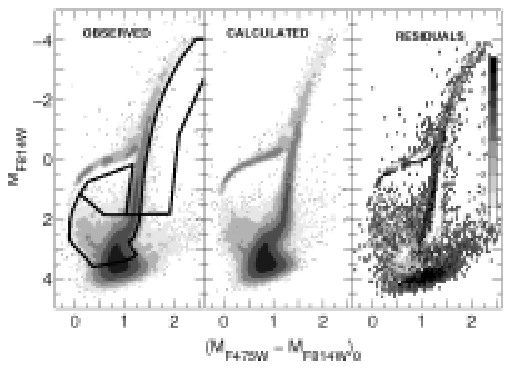}
\caption{Comparison of the observed (left) and best fit (center) CMDs.
The latter was derived by extracting random stars from the model CMD
used to derive the solution, in such a way that each simple population
contributes proportionally to the calculated SFR. The black lines in the
left panel mark the position of the four bundles adopted to derive the
solution. The right panel shows the residual Hess diagram (calculated 
as observed--model), shown in units of Poisson errors. The Figure 
discloses an overall general agreement between the observed and the best-fit
CMDs, particularly in the whole TO region used to calculate the SFH.  The
discrepancy at the faintest magnitude is due to the limiting magnitude of the model
CMD. Note that such a discrepancy does not affect the portion of the CMD included
in the bundles. The differences in the HB morphology are expected, since small differences in the 
adopted mass-loss prescription along the RGB have a big impact on the HB morphology. 
\label{fig:cmdbest}}
\end{figure*}


	\subsection{Comparison of the different solutions}\label{sec:sfh_comparison}

We verified the consistency of the results obtained with the three methods,
observing good agreement as shown in the previous papers. A comparison of
the results obtained using different photometry/library/SFH-code combinations is shown in
Figure \ref{fig:external}, which presents, in the panels $a)$ and $b)$, the
cumulative mass fraction as a function of time for different combinations
of photometry, library, and method. In particular, the panel $a) $compares
the solutions from the DAOPHOT and DOLPHOT photometry derived using the
BaSTI and Padova stellar evolution library. Panel $b)$
shows the solution derived with the three SFH codes, for the DOLPHOT
photometry coupled with the Padova library. The two plots disclose a
generally very good agreement between the different solutions. Panels $c)$
and $d)$ compare the $\psi$(t) calculated with the three SFH algorithms. It has to 
be stressed that the three methods use different IMFs, and the comparison of 
the derived $\psi$(t) should be done carefully. Therefore, panels 
$c)$ and $d)$ compare the three $\psi$(t) before and after applying 
a scaling factor to the MATCH and COLE solutions, in order to take into account
the different IMFs. Note the very good agreement in the main features: the 
position of the main peak at the oldest ages, and the subsequent decreasing
trend. Both the MATCH and COLE methods give some small residual star formation 
between 5 and 8~Gyr ago, which is not present in the IAC-pop solution, but still
the three solutions are consistent within the error bars.

	\subsection{The Tucana SFH}\label{sec:sfhresults}

Similarly to the previous works \citep{monellicetus, hidalgolgs3}, we
discuss in detail the solution based on the IAC method applied
to a model CMD calculated with the BaSTI library and, in particular,
the average of the two solutions based on the two sets of photometry.

We present two different representations of both solutions. Fig.
\ref{fig:tucana3d} shows a 3D view of the best solutions, calculated as
the running average of the 24 individual solutions at the $\chi^2_{\nu\,min}$ 
\citep{hidalgolgs3}. This is a convenient way to have
an immediate, general visualization of  the stellar content and evolution of
a system \citep{hodge89}, since it includes the information on the star
formation rate as a function of both age and metallicity, and the age and
metallicity  distributions as projections of the first. Fig. \ref{fig:tucana3d} 
shows that the bulk of star formation occurred in Tucana
at a very early time and with a limited metallicity range. Only a small
percentage of the star formation inferred in the solution occurred at epochs 
younger than 10~Gyr ago and with metallicity Z $>$ 0.001. Another feature 
that appears in both solutions is a small population of intermediate-age, 
low metallicity stars. We interpret this as a population of blue stragglers,
because the inferred metallicity is in good agreement with that of the oldest 
population in Tucana, and it does not follow the general age-metallicity relation
of the bulk of stars. The same feature was present in the Cetus SFH presented by
\citet{monellicetus}, and was interpreted similarly. We refer to that reference 
for further discussion.


\begin{deluxetable}{lcc}
\tabletypesize{\scriptsize}
\tablewidth{0pt}
\tablecaption{Integrated quantities derived for the Tucana dSph.\label{tab:tab1}}
\tablehead{
\colhead{ } & \colhead{DAOPHOT} & \colhead{DOLPHOT}}
\startdata
$\int\psi(t)dt$ $[10^{6}M_{\odot}]$                    &  (2.52$\pm$0.02)  &  (2.62$\pm$0.03) \\
$<\psi(t)>       [10^{-7} M_{\odot} yr^{-1} pc^{-2}$]  &  (1.25$\pm$0.01)  &  (1.30$\pm$0.01) \\
$<age>           [10^{10}yr]$                          &  (1.15$\pm$0.02)  &  (1.17$\pm$0.01) \\
$<[Fe/H]>        [10^{-4} dex]$                        &  (5.78$\pm$0.83)  &  (5.63$\pm$0.82) 
\enddata
\end{deluxetable}


A complementary view is shown in Fig. \ref{fig:sfh_daodol}, where the star
formation rate, the chemical enrichment law and the cumulative  mass
fraction are plotted as functions of time. The thick continuous and dashed
lines show the DAOPHOT and DOLPHOT solutions, respectively, while the thin
lines represent the corresponding error bars. These have been calculated
following the prescription given in \citet{hidalgolgs3}. Note the good
agreement between the solutions based on the two different sets of photometry. 
Both recover 
the highest peak as a flat feature at the oldest ages, t $>$ 12.5$\,$Gyr ago.
The steepness of the decline is marginally different, and both solutions
give very similar ages at which the star formation completely stopped
in Tucana, around 8--9~Gyr ago. There is also excellent agreement in the small
peak detected at $\sim$ 4~Gyr ago, related to the BS population.
An excellent correspondence is also observed in the derived metallicity laws,
represented in the middle panel. A steady increase of [M/H] is observed, 
from [M/H]$\approx$--1.6 at very early times to [M/H]$\approx$--1.1 at the
end of the star formation activity. After that, the shape of Z(t) is 
necessarily uncertain given the small number of stars from which it is 
derived and, therefore, it is not shown in the plot.

Since the small differences in $\psi(t)$ and in the chemical evolution 
law from the two photometry sets are well within the error bars and because we
do not have any reason to prefer one photometry set over the other, it seems 
reasonable  to adopt the average of the two solutions as the Tucana SFH, and 
the differences between the two as an indication of external errors.
The thick cyan lines in Fig. \ref{fig:sfh_daodol} represent this average,
which we adopt as the final solution for the Tucana SFH.


\begin{figure*}
\epsscale{1.2}
\plotone{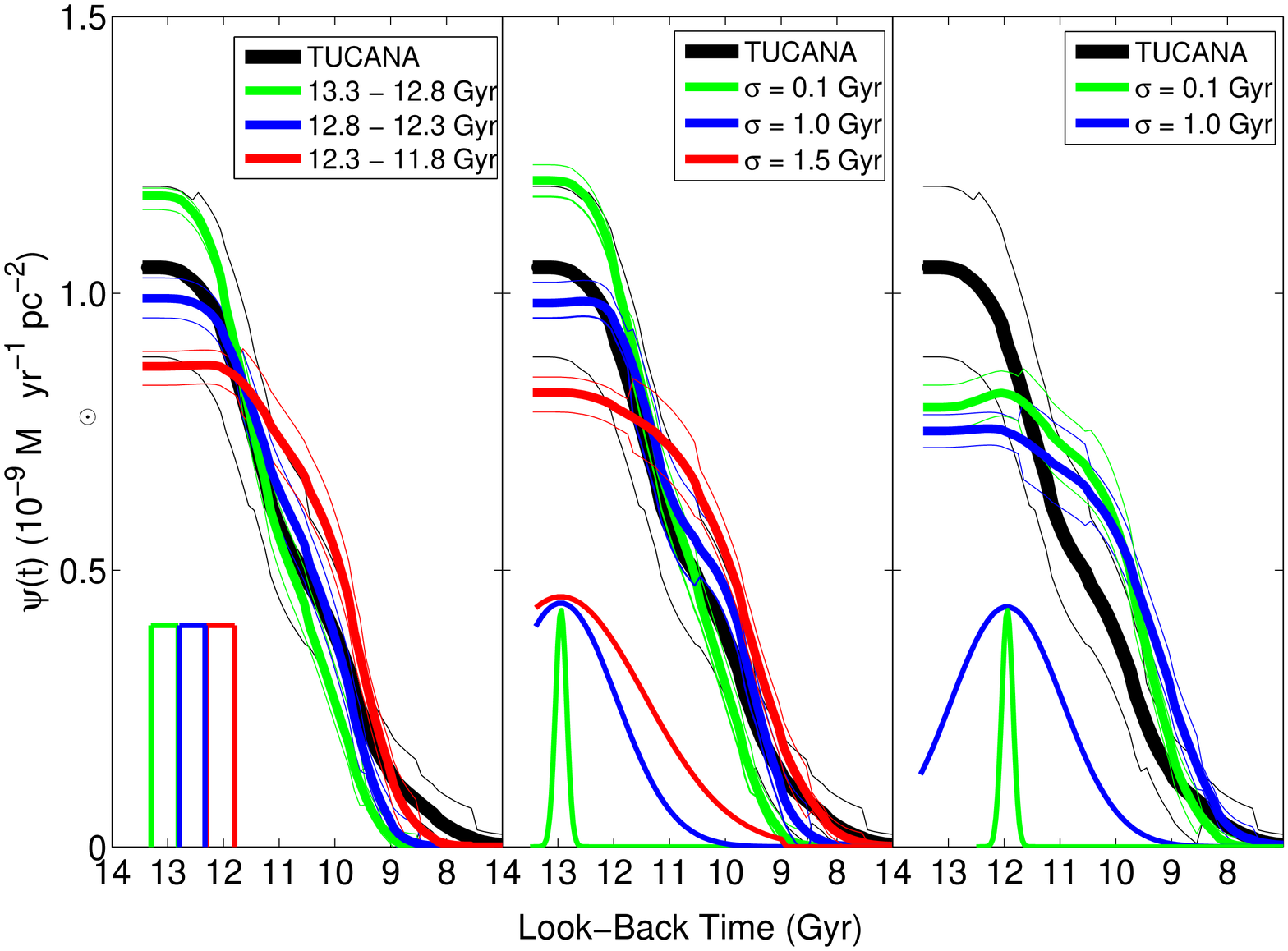}
\caption{Results of the tests performed recovering the SFH of two mock 
populations. The left panel compares the Tucana $\psi$((t)  with that of three
mock bursts at different ages. The central and right panels compare it with
Gaussian profile $\psi$(t) laws of different $\sigma$ and mean age. These comparisons disclose
that the bulk of the star formation in Tucana occurred between 13.5 and 12~Gyr ago,
suggesting that we can not put firm constraints on the effect of the reionization
in this galaxy. Nevertheless, these tests strongly suggest that Tucana could not 
undergo {\it strong\/} star formation at epochs younger than 12~Gyr ago, although
some {\it residual\/} star formation is still compatible with the observed $\psi$(t).
\label{fig:mock}}
\end{figure*}


The upper panel of Fig. \ref{fig:sfh_daodol} discloses that no star
formation occurred in Tucana in the last  $\approx$ 8--9~Gyr. This statement
can be further quantified by quoting the ages when we estimate that 10\%, 50\% 
and 95\% of the stellar mass was formed ($\approx$ 13.2, 12.1 and 9.7~Gyr ago; 
these are represented as vertical lines in the middle panel of Fig. 
\ref{fig:sfh_daodol}). This result demonstrates that Tucana, like
Cetus \citep{monellicetus}, is an outlier of the morphology-density 
relation, because it is a purely old system located at a large distance 
from both the giant galaxies of the LG.

Fig. \ref{fig:cmdbest} presents a comparison between the observed DAOPHOT
CMD and a best-fit CMD (left and central panels), represented as Hess 
diagrams. The bundles used to derive the solution have been superimposed on
the observed CMD. We find satisfactory agreement in all the main evolutionary 
phases. As far as the 
differences in the HB morphology are concerned, it is worth noting that we
have not tried to properly reproduce this observational feature. Due to the
strong dependence of the HB morphology on the mass-loss processes along
the RGB, it is clear that a better agreement between the observational data
and the synthetic CMD could be obtained with a better knowledge of the average mass
loss efficiency and its spread. The right panel shows the residual Hess
diagram, in units of Poisson error, which supports the good agreement 
between the observed and the best fit CMD.

The mean quantities derived (star formation rate, age, and metallicity), 
together with the integral of the $\psi$(t), are summarized in Table 
\ref{tab:tab1}.

	\subsection{The short first episode of star formation in Tucana: 
	are we seeing the effects of reionization?} \label{sec:sfhold}
	
The description presented in the previous section corresponds to the SFH of
Tucana as derived using the algorithms discussed above. This solution is the
convolution of the {\it actual\/} Tucana SFH with some smearing produced by
the combination of observational errors and the limitations of the SFH
recovery method \citep{iacpop}. These are related in part to
intrinsic limitations of the method, such as the small differences in
position in the MSTO of old stars of ages within 2-3~Gyr. In the
following, we will try to constrain further the features of the {\it
actual\/}, underlying Tucana SFH through some tests with mock stellar
populations.


\begin{figure}
\epsscale{2}
\hspace{-3truecm}
\resizebox{15truecm}{10truecm}{\includegraphics[clip=true]{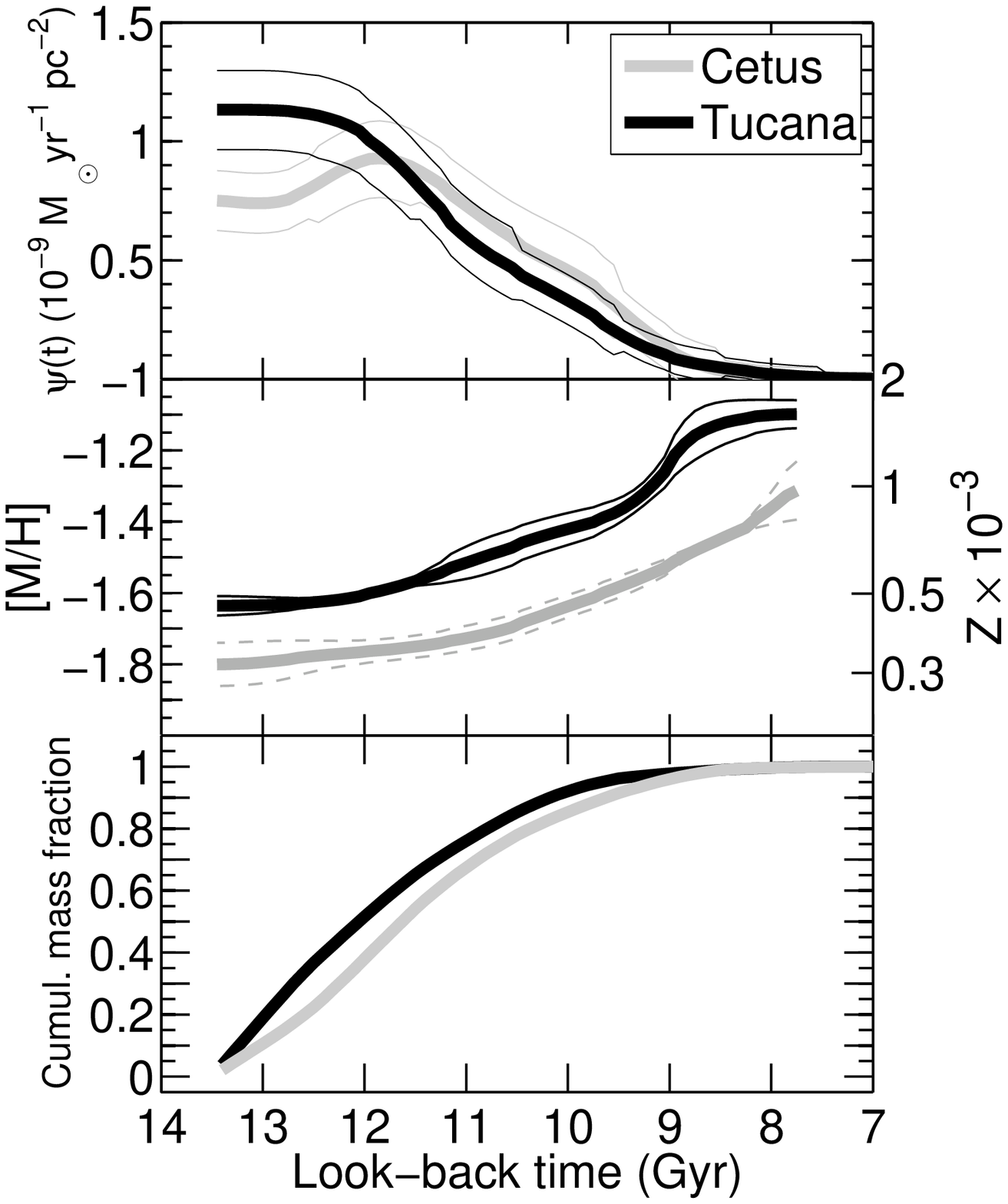}} 
\caption{Comparison of the $\psi$(t) ({\it top}), age-metallicity relations ({\it middle}), 
and cumulative mass functions of Tucana and Cetus. The epoch of peak star 
formation in Tucana is $\sim$ 1.0~Gyr earlier than in Cetus.
\label{fig:cetucana_sfh}}
\end{figure}


In particular, to address the question: "Did cosmic reionization affect the
SFH of Tucana?" we designed a specific test with mock populations. It is
generally assumed that the Universe was completely reionized by $z \approx
6$, equivalent to a look-back time of $\approx$ 12.8~Gyr (Bouwens et al 2007).
Therefore, we created three mock stellar populations with constant $\psi$(t)
in the age ranges 13.3 $<$ t $<$ 12.8~Gyr, 12.8 $<$ t $<$ 12.3~Gyr, and 12.3 
$<$ 11.8~Gyr ago\footnotemark[21]. We also assumed a mean metallicity and a metallicity 
dispersion similar to what is inferred for Tucana. We simulated the observational 
errors of the DAOPHOT photometry, and recovered the SFH of the mock galaxies 
with the same prescription as for the real galaxy. In order to minimize the 
fluctuations associated with the creation of the synthetic CMDs, we repeated 
each test five times. Figure \ref{fig:mock}, left panel, shows the mean of the 
five resulting $\psi$(t) for each mock population, compared with that of Tucana,
shown as a thick black line with the corresponding error bars. Note that the 
fact of repeating the experiment five times reduces the impact of the Poisson errors,
reflected in the smaller error bars of the recovered mock $\psi$(t). The time 
extension of the three input bursts is shown schematically at the bottom of the figure.
The three recovered functions $\psi$(t) show similar general behavior, but the main peak is 
recovered younger and lower, and its width larger, for decreasing age of the input 
$\psi$(t). The comparison with the Tucana SFH leads to two main conclusions: \\
$\bullet$ despite the fact that the best agreement at the oldest epoch is obtained with the 
intermediate-age mock population, the case of the oldest mock population is also compatible, 
within the error bars, with the SFH observed for Tucana. This suggests
that, due to the limited age resolution, we can not draw definitive conclusions
concerning the effect of the reionization on Tucana. In fact, it is equally acceptable that
Tucana started forming stars either before or after the Universe was completely reionized,
more than 12.8~Gyr ago. Moreover, note that if the star formation started before
12.8~Gyr ago, the derived solution may indicate that the star formation 
extends somewhat toward younger epochs. \\
$\bullet$ The youngest mock population is only marginally consistent with the Tucana
solution at all ages, strongly suggesting the the {\itshape bulk} of star formation 
in this galaxy could not occur at epochs younger than $\sim$12~Gyr ago. However, 
it is still possible that some {\itshape residual} star formation was active at 
younger ages.

\footnotetext[21]{Note that the age derived through the SFH recovery depend
on the assumed set of theoretical models. However, it has been shown that the 
libraries adopted in this work give similar ages \citep{gallart05}, and also 
that the ages derived for the Galactic globular clusters are in excellent 
agreement with the age of the Universe from the WMAP experiment \citep{marin09}.
This ensures a meaningful comparison between the solution SFH in the same
cosmological framework.}

Note that the mock populations presented in Figure \ref{fig:mock} are 
characterized by a relatively short episode of star formation, lasting 
0.5~Gyr. Nevertheless, the recovered $\psi$(t) shows a significantly 
longer duration, comparable to that of Tucana. This suggests that the 
actual duration of the main star formation event in Tucana was significantly 
shorter than the calculated one.

To further investigate this point, we simulated more mock populations, with
Gaussian  input $\psi$(t) peaked at 13~Gyr ago and with age dispersion
$\sigma$ = 0.1,  1.0 and 1.5~Gyr. The first two have also been repeated
assuming an age peak at 12~Gyr ago. The Gaussian $\psi$(t) was truncated at
the oldest age we are considering, 13.5~Gyr, i.e., no stars older than 13.5~Gyr
were allowed in the  mock populations. The results are shown in Figure \ref{fig:mock}, 
central and right panels. As in the previous test, we repeated 
each case five times and show the mean recovered $\psi$(t). The comparison in the
central panel shows that the observed peak at the oldest ages is best 
reproduced by the $\sigma$=1.0~Gyr Gaussian profile, but it is still compatible 
with the narrowest one. 

On the other hand, the narrowest Gaussian profile does not properly represent
the shape of the Tucana $\psi$(t), suggesting that the first event was
somewhat longer than this test case. Similarly, the peak of the widest Gaussian 
profile appears too low 
and flat, and is not compatible with the observed one. The comparison
with the Gaussian $\psi$(t) peaked at 12~Gyr ago, shown in the right panel,
indicates that the bulk of the star formation of these mock populations is too
young and is not compatible with the Tucana $\psi$(t). Note that, in 
equivalent tests performed for the Cetus dSph in \citet{monellicetus}, 
a mock Gaussian population peaked at this age was a good representation 
of the observed $\psi$(t). This suggests that, despite the apparent 
similarities of the calculated SFH of these two galaxies, we are able to 
disentangle subtle age differences between the two (see \S \ref{sec:cetucana}).

We stress that the case of Tucana is possibly the most difficult among the LCID 
galaxies. The combination of both intrinsic effects (it is the most distant 
galaxy in our sample, and the oldest one) and observational ones (the shortest
exposure times due to the South Atlantic Anomaly), contribute to increase the complexity
of the problem and the loss of age resolution at the oldest epochs.


\begin{figure*}
\epsscale{1.2}
\plotone{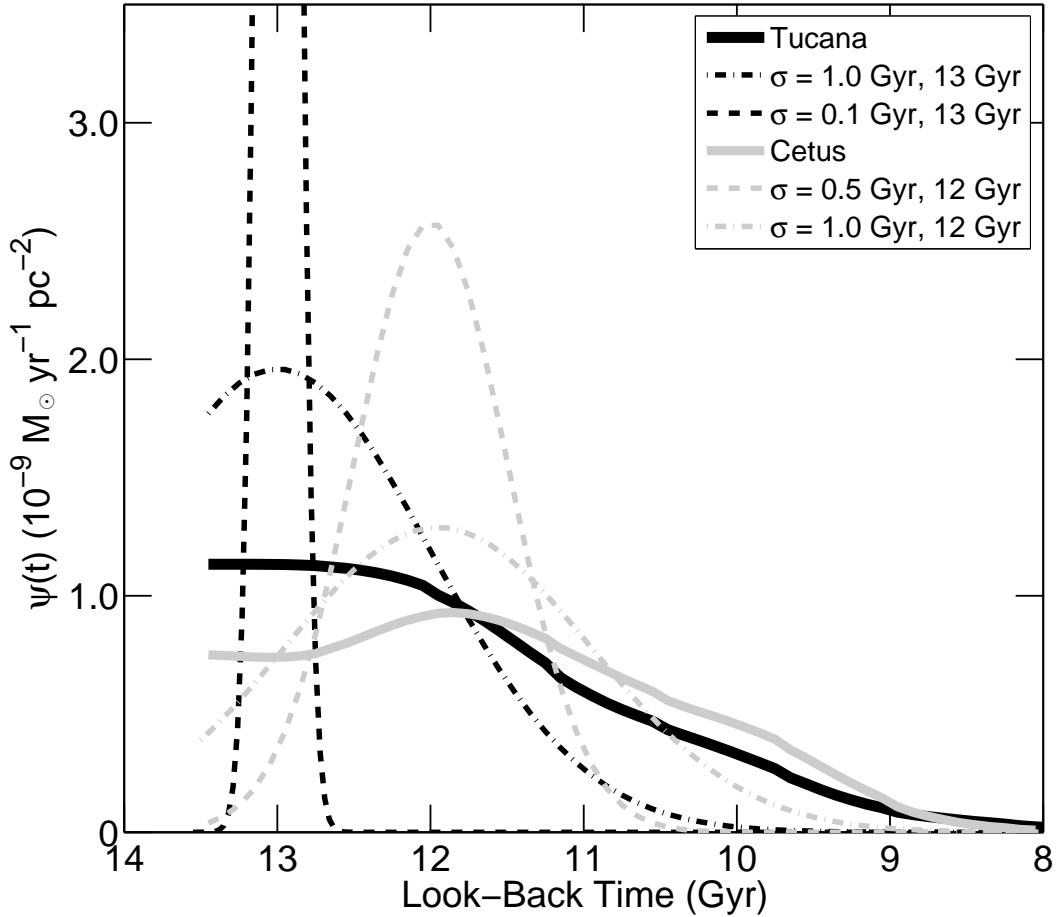}
\caption{Comparison of the Tucana (left) and Cetus (right) $\psi$(t) 
relations with those of Gaussian mock populations giving comparable solutions. 
The inferred underlying SFHs of Tucana and Cetus support the idea of an earlier 
and stronger first episode in Tucana.
\label{fig:actual}}
\end{figure*}


	\subsection{Summary}\label{sec:sfhsummary}

The SFH of Tucana can be summarized as follows. Tucana experienced a very early
and short event of star formation, peaked at ages t $>$ 12.5~Gyr ago. 
We measure a steady decrease in the star formation
rate  until star formation completely stopped $\sim$ 9~Gyr ago. We showed
that the measured duration is overestimated and that the actual event is
compatible with a Gaussian $\psi$(t) with $\sigma$ between $\leq 1.0$~Gyr.
The apparent residual star formation at intermediate ages is interpreted as a
population of blue stragglers. The tests performed with mock populations 
suggest that the epoch of the main episode of star formation is compatible
with the end of the reionization era, and that it can not have occurred later 
than 12~Gyr ago. This is different from what found for Cetus, where the majority
of stars formed at least 1~Gyr later. We will present an in-depth comparison 
of the two galaxies in the next section.

\section{Tucana and Cetus} \label{sec:cetucana}

Tucana and Cetus share the noticeable characteristic of being among the
most isolated dSph galaxies in the LG. Cetus is located at more than 650 Kpc from
both the Milky Way and M31. Tucana is even more distant, at 900 Kpc from 
the Milky Way and 1.3 Mpc from M31. The actual position and orbit are crucial 
to understand interactions that may have occurred in the past. Their receding velocities
with respect of the barycenter of the LG \citep{fraternali09, lewis07} suggest 
that both galaxies may already have experienced at least one passage through the
innermost regions of the LG and, in the case of Cetus, that it is close to
the apocenter of its orbit.

Therefore, a detailed comparison of their properties can be useful to
understand  the origin of the differences and similarities in their SFHs.

	\subsection{Physical parameters}  \label{sec:param}

Tab. \ref{tab:tab2} gives a summary of the intrinsic properties of Cetus
and Tucana. Cetus is the larger galaxy, both in terms of baryonic mass and 
physical size. Tucana presents the mass ($1.81 \times 10^6
M_{\odot}$  baryonic mass, Hidalgo et al. in prep) and size (exponential
scale length $\alpha \sim 120$ pc) typical of the smallest dSphs in the LG, similar
to Sculptor, Leo~II and Ursa Minor. On the other hand, Cetus is similar to
the larger nearby dSphs, with mass and size comparable to Carina or And~II,
with a total baryonic mass of $3.94 \times 10^6 M_{\odot}$ and $\alpha 
\sim 220$ pc. Despite their large distances from both the Milky Way and M31, both Tucana 
and Cetus are similar to the oldest systems like Draco, Ursa Minor or
Sculptor. However, the velocity dispersion estimated for the two galaxies
($\sigma \sim 17 km s^{-1}$) is significantly higher than the measured dispersion
for the nearby dSphs ($\leq 10 km s^{-1}$). Assuming that the circular
velocity of the halo $v_c$ is $v_c = \sqrt(3) \sigma$, and the scaling suggested
for sub-halos by cosmological simulations \citep[see][]{kravtsov10} and confirmed
also by simulations of tidally perturbed satellites with dark matter and stars
\citep{klimentowski07}, one obtains $v_c \sim 29$ km/s for the halo
of Tucana, which would correspond to a virial mass of $\sim 4 \times 10^9 
M_{\odot}$, higher than that of dSph satellites of the MW \citep{lokas09,
strigari10} but comparable
to the mass estimated for their progenitors before they were accreted by
the Milky Way or M31 halos (\citealt[see the reviews by]{mayer10} 
\citealt[and]{kravtsov10}).
Since Tucana and Cetus have been only weakly affected by tides the
estimated mass should indeed be close to their mass before infall into the
primary galaxies.
With such a high halo mass the resulting $M/L$  ratio would be several
hundred in solar units. However, both show hints of rotation, and 
taking this into account would reduce these numbers. Moreover, the faintness 
of the observed stars, together with the low S/N of the available
low-resolution Ca-triplet spectra might also have played a role in overestimating
the velocity dispersion.
It is nonetheless conceivable that $M/L$ is at least as high as 
in the most dark-matter-dominated among the classical dSphs, such as Draco, Ursa
Minor and And~II.

	\subsection{The oldest stars as a benchmark for the reionization effects} \label{sec:cetucana_sfh}

Fig. \ref{fig:cetucana_sfh} compares the derived properties for Tucana (black line)
and Cetus (grey line). The upper panel shows the calculated $\psi$(t)
as the average from the two photometry sets. All the experiments performed  
to test the IAC-pop code \citep{iacpop,
hidalgolgs3, monellicetus}, as well as the tests with mock galaxies
presented in \S \ref{sec:sfhold}, support the reliability of the solutions,
in relation to the determination of the age of the peak star formation
\citep[the typical precision is of the order of 0.5~Gyr, see][]{monellicetus},
including at epochs $>$ 10~Gyr. Moreover, we stress that we are comparing
two SFHs  obtained using the same stellar evolution library, and
the average resulting  from the two photometry sets. This minimizes the
possible systematic effects, giving  more confidence to the estimate of the
{\itshape relative} differences between the main SFH features of the two
galaxies. 

The analysis of the $\psi$(t) shows that the evolution of the two 
galaxies was different. Tucana experienced its strongest level of star 
formation at the earliest epochs, with the initial peak followed
by a steady decrease in the star formation rate. On the other hand, the 
star formation in Cetus \citep{monellicetus} might have started at the 
same epoch, but with the rate increasing until reaching the main peak 
$\sim$ 1.0~Gyr later. Therefore, there is a strong indication that
Cetus formed the majority of its stellar mass significantly after
Tucana, as summarized in Tab. \ref{tab:tab2} with additional information 
such as $t_{50}$ and the mean age.

Note that we have observed most of Tucana's body, while in the case of
Cetus, the very central region of the galaxy (within R$_{gc} \le 0.5 r_c$)
was not contained in our ACS data. This may affect somewhat the resulting
SFH, in view of the stellar population gradients that are present in these
galaxies (which will be discussed in detail in Hidalgo et al.,  in prep.).
However, in this study we found---in agreement with previous works
\citep{gallagher98, harbeck01, io, battaglia06, bernard08, lee09}---that the stellar
population  gradients are in the sense that the population is on average 
younger in the central part.  This implies that the former picture of the
differences between the SFH of Tucana and Cetus would unlikely change (or
if changing, the differences would only increase) if we could compare with
the global SFH of Cetus.

The central panel of Figure \ref{fig:cetucana_sfh} shows the comparison of
the two age-metallicity relations. The thin lines at either side of each
relation  indicate the 1-$\sigma$ dispersion of the metallicity distribution at each
age; the whole metallicity range at each age is much larger. The run of mean
metallicity with age is similar in both galaxies, with the mean metallicity
steadily increasing as a function of time. However, Tucana appears more
metal-rich than Cetus, at fixed age, in contrast with the
metallicity-luminosity relation of dSph galaxies \citep{grebel03}, but still
consistent with the scatter of that relation. Note that the higher mean
metallicity in Tucana does not imply a difference in the metallicity of the
most metal-poor populations, which is the same in both galaxies and
consistent with Z=0.0001. The higher mean metallicity derived for Tucana may
also be connected to the stronger initial event of star formation, which
enriched the medium in a short time scale that we are not able to resolve.

The lower panel of Figure \ref{fig:cetucana_sfh} compares the cumulative mass fraction as
a function of time. Tucana presents a steeper curve, reflecting the discussed stronger
initial activity.

As discussed in  \citet{hidalgolgs3} and in \citet{monellicetus}, the effect
on the derived SFH of our limited age resolution at old ages, is mainly to
broaden the main features, and to artificially extend the limits of the
epoch(s) of star formation. The subtle differences between the Tucana and
Cetus SFHs may be best visualized in terms of the possible {\it true,
underlying} SFHs that we have discussed for both galaxies, using tests with
mock stellar populations. Figure \ref{fig:actual} summarizes the comparison of
the $\psi$(t) retrieved for Tucana and Cetus with those of the compatible
Gaussian mock populations. Despite being only qualitative, this plot clearly
discloses that the star formation in Tucana is expected to have been
significantly stronger at the oldest ages than in the case of Cetus and
that, since the narrowest events seem to be compatible with the
SFH of both galaxies, even the main star formation event may have occurred
at almost non-overlapping epochs. Despite the uncertainties possibly
affecting the {\itshape absolute}  age scale, and the ambiguity concerning
the duration of the main event in Tucana  highlighted in \S \ref{sec:sfhold}, 
the {\itshape relative} difference of 1~Gyr calculated between the inferred
peak of the $\psi$(t) of Tucana and that of Cetus strongly reinforces the conclusions 
presented in \citet{monellicetus}, where we argued that the reionization had 
limited, if any, effects on the star formation of Cetus. This is expected
if the halo masses of these dwarfs are $> 10^8 M_{\odot}$ \citep{susa04,
kravtsov10}, which can be easily argued even if their velocity 
dispersion is somewhat overestimated. We can rule 
out the possibility that Cetus experienced the majority of its star formation at
epochs earlier than the end of the reionization. This conclusion is solid
and is further sustained by the consistent results obtained using different
photometry sets, stellar evolution libraries and SFH codes adopted for the
LCID project.


\begin{deluxetable*}{l|r|r}
\tabletypesize{\scriptsize}
\tablewidth{0pt}
\tablecaption{Estimates of the basic parameters for the two dSph galaxies Tucana and Cetus }
\tablehead{
\colhead{parameter}      &  \colhead{Tucana}                                    & \colhead{Cetus}     } 
\startdata
D [Kpc]                  &  $   887\pm50 \tablenotemark{d}             $  &  $  780\pm43 \tablenotemark{d}             $   \\
(m-M)$_0$                &  $  24.74\pm0.12\tablenotemark{d}           $  &  $  24.46\pm0.12\tablenotemark{d}          $   \\
M$_V$  [mag]             &  $ -9.55\pm0.27\tablenotemark{e}	       $  &  $ -10.1\tablenotemark{a}; -11.3\tablenotemark{b}   $  \\ 
L$_V$  [L$\odot$]        &  $  5.2\times10^5 \tablenotemark{e}         $  &  $  8.7\times10^5 \tablenotemark{a}; 2.6\times10^6 \tablenotemark{b}     $  \\ 
M/L  [M$\odot$/L$\odot$] &  $  105^{\rm +95}_{\rm -49}\tablenotemark{c}$  &  $  \simeq 70 \tablenotemark{f}            $  \\ 
Total Mass [M$\odot$]    &  $  \simeq 6\times10^7\tablenotemark{c}     $  &  $   \simeq 6\times10^7 \tablenotemark{f}  $  \\
Stellar Mass [M$\odot$]  &  $   1.81 \times10^6                        $  &  $    3.94 \times10^6                      $  \\
$\alpha [\arcmin (pc)]$  &  $  0.48\pm0.08 (124) \tablenotemark{e}; 0.51 (132) \tablenotemark{g} $  & $  1.0\pm XX (230) \tablenotemark{g}; 1.59\pm0.05 \tablenotemark{b}  $  \\ 
Rc   [$\arcmin (pc)$]    &  $  0.59 (152)     \tablenotemark{d}; 0.66\pm0.1 (170\pm25)\tablenotemark{e}	$  & $  1.5\pm0.1 (340)\tablenotemark{a}; 1.3\pm0.1 (295)\tablenotemark{b}$  \\
Rt  [$\arcmin (pc)$]     &  $  3.45 (890)     \tablenotemark{e}	       $  &  $  4.8\pm0.2  (1089\pm45)\tablenotemark{a}; 32\pm6.5   (7261\pm1475) \tablenotemark{b}$  \\
c                        &  $  0.72	      \tablenotemark{e}	       $  &  $  0.5\tablenotemark{a};  1.4\tablenotemark{b}     $  \\
$\mu_{o,V}$ (obs)[mag]   &  $  25.05\pm 0.06  \tablenotemark{e}	       $  &  $  25.0 \tablenotemark{b}		       $  \\
$v_{hel}$ [km/s]         &  $  +194.0\pm4.3   \tablenotemark{c}	       $  &  $  -87\pm2  \tablenotemark{f}             $  \\
$v_{GSR}$ [km/s]         &  $  +98.9          \tablenotemark{c}	       $  &  $  -25  \tablenotemark{f}                 $  \\
$v_{LGSR}$ [km/s]        &  $  +73.3          \tablenotemark{c}	       $  &  $  +14 \tablenotemark{f}	               $  \\
$\sigma$ [km/s]          &  $  17.4_{\rm +4.1}^{\rm -3.1} \tablenotemark{c} $ & $  17 \pm 2 (14\div8)\tablenotemark{f} $  \\
vrot [km/s]              &  $  \simeq 16      \tablenotemark{c}	       $  &  $  7.7 \pm 1.2\tablenotemark{f}	       $  \\
$<$age$>$ [Gyr]          & $    11.6 \pm 0.1 \tablenotemark{h}         $ &  $   11.3  \pm 0.2     \tablenotemark{g}    $  \\
T$_{10}$  [Gyr]          & $    13.2         \tablenotemark{h}	       $ &  $   13.1      \tablenotemark{g}            $  \\
T$_{50}$  [Gyr]          & $    12.1	     \tablenotemark{h}	       $ &  $   11.6      \tablenotemark{g} 	       $  \\
T$_{95}$  [Gyr]          & $     9.7	     \tablenotemark{h}         $ &  $	 9.6      \tablenotemark{g} 	       $  \\
                         & $                                           $ &  $                                          $  \\
\enddata
\tablenotetext{a}{\citealt{whiting99};}
\tablenotetext{b}{\citealt{mcconnachie06};}
\tablenotetext{c}{\citealt{fraternali09};}
\tablenotetext{d}{\citealt{bernard09};}
\tablenotetext{e}{\citealt{saviane96};}
\tablenotetext{f}{\citealt{lewis07};}
\tablenotetext{g}{\citealt{monellicetus};}
\tablenotetext{h}{this work.}
\label{tab:tab2}
\end{deluxetable*}


	\subsection{Morphology of the HB and RR Lyrae stars properties}\label{sec:cetucana_cmd}

Is it possible to link the resulting SFHs with other independent
evidence? A comparison of the Tucana and Cetus CMDs is shown in Fig.
\ref{fig:cetucana_cmd}. At a first glance, the striking similarities seen
in the shape of the TO region and in the slope and color width of the RGB
suggest a similar age range and spread of metal content.

An obvious difference between the two shows up in the morphology of the HB.
The HB in Tucana presents a complex morphology, with both the blue and the red
side well populated, the red side also presenting a split into two parallel
sequences. Cetus presents a redder HB morphology \citep{sarajedini02,
monellicetus}, but nonetheless the blue side is populated, even if by a 
many fewer stars. The color of the bluest Cetus HB stars
is the same as for Tucana. 

Therefore, the color limits of the HB are similar in the two galaxies, at 
both extremes. If we assume that, in complex systems like these, the prolonged 
star formation and the associated chemical evolution drive such a color spread, 
then we can expect that the blue side is populated by the oldest, most metal-poor
stars. Therefore, the higher number of stars in this region of the Tucana 
CMD suggests a much stronger star formation event at older epochs and/or lower
metallicities than for 
Cetus. However, the presence of a few blue HB stars in the latter supports 
the idea that some amount of very old population s present in both. These considerations are in 
excellent agreement with the derived SFHs (see Figures \ref{fig:mock} and
\ref{fig:actual}).

The same scenario is also nicely supported by the properties of the RR Lyrae
stars. \citet{bernard08} showed the presence of two sub-populations of RR
Lyrae stars in Tucana, characterized by different mean luminosities, different mean
periods, different slopes of the period-amplitude relation, and different radial distributions. 
The interpretation of this evidence is that the two sub-populations have
different properties in terms of age and/or metallicity. Note that the
fainter sample, presumably the younger and more metal-rich, is also the
more centrally concentrated.  A similar dichotomy was not detected in Cetus,
where the sample of RR Lyrae presents homogeneous properties.

The presence of two sub-populations of old stars in Tucana, supported by both
the split in the red HB and the properties of the RR Lyrae stars, implies
that that the first generation is numerous enough to show up clearly and that
Tucana was able to produce a second generation of stars with distinct chemical 
properties due to the self-enrichment from the previous event. The time
resolution achievable with the present data is not enough to distinguish
an episodic from a continuous $\psi$(t). Nevertheless, the chemical
enrichment seems to have occurred on very short time scale. The fact that 
the mean metallicity is relatively high even at the oldest epochs (see Figure
\ref{fig:cetucana_sfh}, central panel), indicates that the
enrichment occurred very early. Therefore, the derived SFH, dominated by
a single event, is still compatible with two populations of old stars
identified through other evidence.

	\subsection{The RGB bump}\label{sec:cetucana_rgb}

The RGB bump is a well-defined feature in the CMD of both galaxies 
\citep[see Figure \ref{fig:cetucana_cmd} and ][]{monelli10b}.  In
particular, Tucana presents two bumps, a bluer and brighter one, well
separated from a second which is redder and fainter. Cetus only shows one well
populated bump, located on the red side of the RGB. In \citet{monelli10b},
we discussed how these observed features can be interpreted using the SFHs of
the two galaxies, and we put constraints on the age and the metallicity of
the populations which produce them. More metal-poor stars (Z$<$0.0005)
produce the brighter bump in Tucana, while the fainter, which is also more
populated, consists of slightly more metal-rich stars (0.0005 $<$ Z $<$
0.001). In Cetus, the dominant  population (0.0003 $<$ Z $<$ 0.0006)
produces the single bump. The RGB bumps, therefore, also reflect the
previously discussed differences in the Tucana and Cetus SFHs, with an older, more metal
poor population in Tucana numerous enough to show in a separate RGB bump in
the blue part of the RGB.

	\subsection{Summary}\label{sec:cetucana_rgb}

All these pieces of information give a coherent picture of the evolution of
Cetus and Tucana and of the origin of the subtle differences we
have detected. The SFH of Tucana is compatible with a very old and brief
event. The presence of two sub-populations of old stars in Tucana, supported
both by the split in the red HB, the properties of the RR Lyrae stars and
the double RGB-bump, implies that Tucana was able to produce a second
generation of stars with distinct chemical properties due to 
self-enrichment from the first stars. The important fraction of very
metal-poor stars in Tucana indicates a high efficiency of star
formation at the earliest stages. We cannot distinguish,  with the 
present data, between two discrete events or a continuous process.

On the other hand, Cetus started forming stars most likely at the same epoch, 
as indicated by the bluest HB stars. However, Cetus seems to have had a more gradual
evolution, reaching its peak star formation 1~Gyr after Tucana, and producing 
a dominant population with more homogeneous metallicity. It contains a minority 
of old, metal poor stars, like the oldest stars in Tucana, and a majority of 
slightly younger stars (and possibly somewhat metal enriched with respect to the
oldest). The different mean metallicities of Cetus and Tucana can be possibly
explained by the intense initial episode of star formation that Tucana
experienced at older epochs, that enriched the medium causing the higher
metallicity of the subsequent generation of stars.

According to the most widely accepted cosmological framework, the age of the
Universe is $\sim$13.7~Gyr, and the best estimate of the reionization epoch is
at $z \sim 10.9 \pm 1.4$ \citep{komatsu09}. Therefore, the most important 
conclusion we can draw from the comparison of the SFH of Cetus and Tucana is 
that the vast majority of the stars in Cetus were born after the end of the
reionization, currently assumed at $z \sim 6$. This confirms that the reionization 
can not have been the dominant mechanism causing the end of star formation in Cetus
\citep{monellicetus}. On the other hand, the timescale of the star formation 
in Tucana suggests that the reionization may have played an important role
in this galaxy.

\section{Discussion}\label{sec:discussion}

The SFH of the Tucana isolated dSph galaxy presented in this paper shows
that the star formation was very active at very early epochs ($>$ 12.5~Gyr
ago) and over a short time scale (compatible with a Gaussian $\psi$(t) with
$\sigma$ in the range t $\leq$ 1.0~Gyr). The tests we performed with
mock galaxies show that the present data do not allow enough time resolution
to firmly correlate the epoch of the reionization with the end of the star formation 
in Tucana. The comparison with Cetus discloses that the latter undoubtedly 
experienced delayed activity, as compared with Tucana, and the vast majority 
of its stars were formed  significantly later than the end of the reionization 
era. This would be compatible with the higher mass of Cetus, and with the fact 
that lower mass systems are expected to form earlier \citep{kravtsov10}. However, 
it is particularly intriguing that LGS3 is found to be less massive than Tucana, 
but nevertheless it formed 80\%  of its stars well after the end of the 
reionization, with a peak star  formation at the same age as in Cetus 
\citep{hidalgolgs3}.

Given this, if we assume that the end of the star formation was not predominantly
induced by the reionization, what was the cause? In the following, we explore 
possible alternative mechanisms, both internal and external, that might have
affected the evolution of Tucana.

	\subsection{The effect of SN explosions}\label{sec:sn}

We will explore here the possibility that the star formation in Tucana was stopped by
the removal of the gas by purely internal processes, such as the explosion of supernovae
(SNe). Given the adopted IMF (assumed from \citealt{kroupa01}), coupled with
the accurate estimate of the total mass transformed into stars in the 
galaxy  (obtained as the
integral of the star formation in the main event, reported in Table
\ref{tab:tab2}), it is possible to calculate the expected total number of
type II SNe originating from the explosion of massive stars. At the
metallicity typical of the Tucana population, we can assume that SNe II
originate from stars with masses larger than $\approx 6.5 M_{\odot}$.
Integrating the IMF, and assuming 3.2 $\times 10^6 M_{\odot}$ to be the mass of stars formed
as derived from the SFH, we estimate a total number of $\sim$54,300
SNe II events. This total number does not depend on the duration of the burst,
but the number of SNs per unit time increases for decreasing duration
of the star-formation event. For the two cases examined in \S
\ref{sec:sfhold} we can assume a duration equal to the FWHM of the input
Gaussian: 0.235~Gyr and 1.67~Gyr for the $\sigma = $0.1 and 1.0~Gyr cases. Note
that, in the latter case, the  assumed duration is smaller than the
theoretical value of 2.35~Gyr because we consider only stars younger than
13.5~Gyr. Assuming that each SN releases 10$^{51} erg$, this implies a
total luminosity of $7.33 \times 10^{39} erg s^{-1}$ and $1.03 \times
10^{39} erg s^{-1}$, for the 0.1 and 1.0~Gyr events, 
respectively\footnotemark[22]. \footnotetext[2]{Note that this can be
considered a conservative upper limit, since the total number of expected SN
critically depends on the minimum stellar mass assumed to produce a SN
event. This, in turn, strongly depends on the overshooting and the mass 
loss experienced by the stars during their evolution: the effect of mass 
loss is to increase the mass  of stars ending as a SN, therefore decreasing 
the number of events.}
If we  compare these numbers with the model proposed by \citet[][see their
Figure 1]{maclow99}, we can see that in both cases Tucana is in the so-called 
blowout regime, meaning that part of the gas must be lost due to the
wind produced by SNe. 
This gives a general indication that gas loss, and especially 
the loss of metals directly connected with the SN winds, are efficient 
mechanisms in small systems like Tucana \citep{maclow99, ferrara00, sawala10}.

However, we have shown strong evidence that two stellar populations
with  different properties formed in Tucana on a short time scale, and that
the younger population was enriched by the yield of the previous one. This
has two important implications: $i)$ Tucana was able to retain a sizable
fraction of its gas after a first episode, and $ii)$ the time between the
two episodes of star formation was long enough to enrich and mix the medium. Given
these two results, it is possible to estimate both the minimum time scale 
required to start enriching the medium, and amount of yield that has been 
produced by the first generation and recycled by the second one, and thus to
estimate the amount of metals lost in an independent way. 

In the case of SNe II, the typical time scale is extremely short, of the order 
of few Myr. In the case of SNe Ia,
this is related to the evolution of intermediate-mass stars in binary
systems, which ends up with  a degenerate CO core of mass insufficient for quiet
ignition of the C burning. The mass  limit of the star that produces a degenerate CO
core has been shown to depend on the metallicity of the star \citep{cassisi93}. 
For the assumed peak metallicity of the oldest population, the star with largest
initial mass capable of producing a degenerate CO core is expected to be $\approx
6.5 M_{\odot}$ star. The evolutionary time of such stars fixes a minimum
time scale for the creation of SNe Ia of $\sim$ 60 Myr. This suggests
that, even in the scenario of the shortest Gaussian profile burst, 
SNe Ia were capable of contributing to the enrichment of the interstellar medium of
Tucana.

Second, it is possible to estimate the number of SNe required to produce the
iron that enriched the more metal-rich population in Tucana. The morphology 
of the HB suggests that the two populations of old stars are compatible 
with mean metallicities Z = 0.0002$\pm$0.0001 and Z = 0.0006$\pm$0.0002, 
according to the comparison with theoretical ZAHBs and the SFH results. Based 
on the star counts on the HB, where the two populations can be more easily
split, we find that each of the two episodes formed $\sim$50\% of the stars.
Assuming the above metallicities, and considering the total mass of stars 
formed, we can calculate the total mass of the metals contained in the younger 
population. In particular, assuming that each episode formed 1.6 $\times 10^6 
M_{\odot}$, it turns out that 640 $M_{\odot}$ of heavy elements are recycled 
by the second generation. Assuming the Solar mixture from \citet{grevesse93}, 
the amount of iron expected is $\approx 46 M_{\odot}$. If we assume
that only SNe II are involved, using the iron mass production rate from
\citet{nomoto97b}, we estimate that $\sim$ 300 SNe II are enough to produce
this amount of iron. This estimate was derived assuming the iron mass production 
of a 13 $M_{\odot}$ star. This value might change by a factor of 2--3 with
different assumptions. We can now compare this number with the
total expected number of SNe II derived before. If we take into account that
stars more massive than $\sim 25 M_{\odot}$ do not enrich the medium,
because they end up as neutron stars or black holes, we derive that, from a
total of $\sim$23,540 SNe II releasing iron during the first star formation
event, only $\approx$1.3\% of the iron mass is recycled, while the vast
majority is lost, in agreement with the model by \citet{maclow99}. On the
other extreme, since SNe Ia are more efficient producers of iron
\citep{nomoto97a}, we estimate that only $\sim $70 SNe Ia are required to 
produce the iron present in the younger population and that therefore an
even larger fraction of gas is lost.

	\subsection{The effect of the environment}\label{sec:int}

The SFH presented in this paper proves that Tucana, like Cetus,
is an outlier in the morphology-density relation. Despite its large distance 
from both the Milky Way and M31, Tucana is a purely old galaxy, similar to 
the oldest satellites of the Milky Way like Draco, Ursa Minor, or Sculptor,
with which it seems to share also a very high $M/L$. This poses interesting 
questions on the effectiveness of the environment in shaping the SFH. Why is a 
galaxy that spent most of its life in isolation so similar to galaxies of 
similar mass which are close satellites? The idea that close encounters with
giant galaxies can influence the evolution of dwarf galaxies is a 
well-established  prediction of models \citep{mayer01, mayer10, lokas10}. 
However, at least 2--3 orbits on relatively small pericenters ($< 50$ kpc) 
are typically required to completely transform an initially gas-rich disky-like 
dwarf into a dSph, leaving a nearly isotropic stellar component supported by 
stellar velocity dispersion rather than rotation \citep{klimentowski09, lokas09}. 
A single passage on a relatively wide orbit (for an orbit with a 1:10 ratio 
between apocenter and pericenter, which is slightly larger than the mean but 
still common for sub-halos in cosmological LCDM simulations---see, e.g.,
\citealt{diemand07} and \citealt{klimentowski10} the pericenter distance of Tucana 
must have been at $R > 50$ kpc) would produce a remnant retaining significant 
rotation and deviations from sphericity \citep{kazantzidis10}. While the residual
rotation in Tucana argues in favor of the latter possibility, its structure and 
kinematics are not firm enough yet to allow drawing conclusions of this sort.

However, based on actual knowledge of the orbit of Tucana, 
it is concluded that it may have experienced at most one close encounter with
a LG giant galaxy. In fact, the velocity of Tucana with respect to the barycenter
of the LG, $v_{LGSR} = +73.3 km s^{-1}$ \citep{fraternali09}, suggests that a
close encounter in the densest LG regions was possible, and in particular
the velocity with respect to the Milky Way, $v_{GSR}  = +98.9 km s^{-1}$, is
compatible with an encounter that occurred at $t_{enc} \leq $10~Gyr ago
\citep{fraternali09}. This is a somewhat later than the main star-formation
epoch identified in this paper, which appears to have
ended between 12.5 and 11.5~Gyr ago, but two events could have been
nearly simultaneous, given the large uncertainties. A single passage at a distance of tens of kpc
could be enough to strip most of the interstellar medium of the galaxy
by ram pressure \citep{mayer07}, provided that the coronal gas density around the 
proto-Milky Way was as high as suggested by observational constraints on its present 
state ($\sim 10^{-4} atoms/cm^{3}$). Heating by supernova feedback and by the 
cosmic ionizing background, which remained high down to $z\approx2$,
would have rendered the gas more diffuse and pressure-supported, facilitating
stripping by the combined action of ram pressure and tides \citep{mayer06, 
mayer10}. However, it should be taken
into  account that at such early epochs the size of the Milky Way and M31 was
significantly  smaller, so the effects of a close encounter might not
have been as significant. Also, the structure of the
coronal gas at such high redshift, when galaxies are undergoing rapid gas
accretion via cold flows \citep{dekel09, agertz09}, is unclear; while in
principle gas densities around the primary could be even higher than at present
as the filamentary flows are made of cold relatively dense gas, the halo
might not be filled with gas everywhere as in a smooth corona.

Moreover, if it turns out that Tucana has little or no residual rotation 
a complete transformation via tidal perturbations after only one pericenter 
passage would require quite specific conditions, based on gravitational 
resonances occurring during the interaction \citep{donghia09}. Another 
possibility is that Tucana was pushed to its highly eccentric orbit as 
a result of a three-body interaction \citep{sales07}, and thus could have 
had more than one close encounter with one of the primary galaxies. Finally, 
in cosmological simulations a few sub-halos undergo interactions and mergers
with other sub-halos before being accreted by the primaries \citep{kravtsov04,
klimentowski10}, which could also produce a dSph out of two interacting
small disky dwarfs well outside the virial radius of the primary.

On the other hand, the evidence that the properties of nearby and isolated 
dSphs present important similarities might indicate that interactions are 
not {\itshape the} dominant mechanism shaping these galaxies. For example,
the supernova explosions from the first generation of stars contributed to 
the turbulence of the gas. This would have made the gas easier to strip \citep{murakami99}
via ram pressure. Therefore, internal (SN), external (interactions, gas 
stripping, ram pressure) or global (reionization) mechanisms, can be at 
play, and the combination of these might have shaped the morphology of some 
dSphs even before accretion into the satellite system of a large galaxy \citep{sawala10}.
An analysis based on homogeneous SFHs derived for both the Milky Way and
M31 satellites galaxies, together with reliable knowledge of their orbits,
would be needed for a meaningful comparison of their properties  with those 
of Tucana and Cetus, in order to gain some insight into the role of the
environment in the evolution of small systems.

	\subsection{Tucana as a high-$z$ galaxy}\label{sec:highz}

The range of estimated burst duration for Tucana ($\sigma \leq 1.0$~Gyr),
allows us to infer the corresponding 
actual peak intensity of the initial burst: assuming $\sigma \geq 0.1$~Gyr, this ranges from  0.3 to 1.3
$\times 10^{-8} M_{\odot} pc^{-2} yr^{-1}$. Note that this is the average 
value, as calculated over the whole area covered, from the center to $\sim$ 
3 core radii. Preliminary analysis of the $\psi$(t) as a function of radius
(Hidalgo et al. in prep.) shows that the star formation rate in the innermost
regions is $\approx$ 4 times higher than the average value. This means that
the SFR of Tucana, during the  first stages of its evolution, was comparable
to that observed for present day Blue  Compact Dwarf galaxies
\citep[see][and references therein]{annibali03}.

Despite the fact that we can not put a stringent constraint on the end of the 
star formation in Tucana, a solid result of our analysis is that Tucana was 
efficiently forming stars at epochs at least as old as $z \sim 7$. In various 
recent works \citep{bouwens10a, bouwens10b, gonzalez10, labbe10, oesch10}, 
galaxies at similar redshift have been identified and
characterized for the first time. However, all the galaxies so far detected have masses and
luminosity at least two orders of magnitude larger than Tucana
\citep{gonzalez10}.

\section{Summary and conclusions}\label{sec:summa}

Deep HST/ACS data have allowed us to derive the lifetime SFH of the Tucana dwarf
galaxy, one of the most isolated dSphs in the LG. We have shown that Tucana
experienced a strong event of star formation at the oldest possible age, $>
12.5$~Gyr ago. After the first initial peak, the measured intensity of the
star formation steadily decreased until stopping $\approx 9$~Gyr ago. The
tests we performed with mock stellar populations disclose the broadening
effect of the observational errors. We find that the actual underlying SFH
is compatible with an episode of short duration, in the range $\sigma \leq $
1.0~Gyr if we assume a Gaussian profile $\psi$(t) peaked 13~Gyr ago. Our attempt to put
firm constraints on the age limits of the main event of star formation are
hampered by the limited time resolution, and thus we are not able to clearly
answer the question of whether the reionization was decisive in ending Tucana's
star formation activity.  

We explored alternative mechanisms (both external and internal) that may
have shaped Tucana's evolution. On one hand, current measurements of its
radial velocity do not rule out the possibility that Tucana may have traversed the inner
regions of the Local Group once. Therefore, a morphological transformation
linked to a close interaction with a larger Local Group member at a time
consistent with the end of its star formation cannot be ruled out. On the
other hand, we explored the possibility that the feedback from supernova
explosions might have been responsible for the gas loss. We used two
different arguments to conclude that gas loss connected to supernova
events must indeed have been very important in Tucana's early evolution: first,
comparison of the total amount of energy released by the SNe expected in
the early evolution of Tucana with the model by  Mc~Low \& Ferrara (1999)
shows that Tucana could be in the blowout region. Second, investigating the 
chemical difference between 
two main sub-populations of distinct metallicity present in Tucana, we
found evidence that the vast majority of the metals produced by the SNe
must have been lost by Tucana to the inter-galactic medium. 

We devoted a particular effort to compare the properties of Tucana and
Cetus, the two isolated dSphs analyzed consistently in the LCID project, 
that also share the important characteristic of being the most isolated dSphs
in the LG. Both are (at first approximation) as old as the oldest Milky Way
satellites, such as Draco, UMi or Sculptor, with no traces of star
formation younger than $\simeq$ 9~Gyr. The fact that they do not follow
the morphology-density relation that has been observed in Milky Way dSph
satellites poses interesting  questions concerning the effectiveness of the
environment in shaping the SFHs of dwarf galaxies. This, gives
some support to models such as the one recently published by \citet{sawala10}
in which internal mechanisms such as SNe, enhanced by the effects of
cosmic reionization, are able to reproduce the main characteristics of dSph
galaxies without having to invoke strong environmental effects. 
Still, new clues will come from a better understanding of the structure and
kinematics of the stellar component of Tucana, and by comparing such
properties in detail with those of the classical dSph satellites that were 
clearly affected by the environment.

Despite the obvious similarities in the CMDs and SFHs of Cetus and Tucana, 
we also demonstrated important differences in their early evolution. We
have shown that the formation time of the bulk of the stellar populations
in Cetus is clearly delayed compared to Tucana. This clearly
appears in the derived SFHs, and other independent indicators support the
same conclusion: the morphology of the HB, the  properties of the RR Lyrae
variable stars \citep{bernard08, bernard09}, and the characteristics of the
RGB bump \citep{monelli10b}. The most important conclusion we can draw from
this comparison is that it strongly reinforces the conclusions of 
\citet{monellicetus}, in particular that the vast majority of the stars in
Cetus were formed well after the end of the reionization epoch, therefore
suggesting that the end of the star formation in Cetus was not predominantly
caused by it. This has important implications for state-of-the-art
models on the effects of reionization in the early SFH of dwarf galaxies.

\acknowledgments

{\it Facilities:} \facility{HST (ACS)}.

Support for this work was provided by NASA through grant GO-10515
from the Space Telescope Science Institute, which is operated by
AURA, Inc., under NASA contract NAS5-26555, the IAC (grant 310394), the
Education and Science Ministry of Spain (grants AYA2004-06343 and
AYA2007-3E3507).
This research has made use of NASA's Astrophysics Data System
Bibliographic Services and the NASA/IPAC Extragalactic Database
(NED), which is operated by the Jet Propulsion Laboratory, California
Institute of Technology, under contract with the National Aeronautics
and Space Administration.


\end{document}